\documentclass[11pt,a4paper]{article}
\usepackage{jheppub}
\usepackage{amsmath,amssymb,graphicx,epstopdf,enumerate,amsfonts,booktabs,color}
\usepackage{float,placeins}
\providecommand{\U}[1]{\protect\rule{.1in}{.1in}}
\newcommand*\rfrac[2]{{}^{#1}\!/_{#2}}
\affiliation[a]{Department of Electrophysics, National Chiao Tung University, Hsinchu, ROC}
\affiliation[b]{Physics Division, National Center for Theoretical Sciences, Hsinchu, ROC}
\affiliation[c]{Department of Physics, Chung Yuan Christian University, Taoyuan City, ROC}
\affiliation[d]{School of physics, University of Chinese Academy of Sciences, Beijing 100049, China}
\affiliation[e]{Kavli Insititute for Theoretical Sciences, University of Chinese Academy of Sciences, Beijing 100049, China}
\emailAdd{sr755332@gmail.com}
\emailAdd{yiyang@mail.nctu.edu.tw}
\emailAdd{phy.pro.phy@gmail.com}
\abstract{We study the gravitational wave spectrum from the confinement-deconfinement phase transition in the holographic QCD models for both light quark and heavy quark. We obtain the gravitational wave spectrum in the deflagration, detonation and runaway cases for different chemical potentials and duration of the phase transition. We find that the gravitational wave spectrum could be observed by EPTA/IPTA and eLISA for large chemical potential and short duartion of phase transtions.}
\begin{document}

\title{Imprints of Early Universe on Gravitational Waves from First-Order Phase Transition in QCD}
\author{Meng-Wei Li${^{a}}$, Yi Yang${^{a,b}}$ and Pei-Hung Yuan${^{a,c,d,e}}$}
\maketitle

\setcounter{equation}{0}
\renewcommand{\theequation}{\arabic{section}.\arabic{equation}}

\section{Introduction}
The idea of Gravitational wave (GW) was first proposed by Albert Einstein in 1916 \cite{Einstein1916,Einstein1918}, and have been developing our understanding by Josh Goldberg in 1957, especially how a binary star system generates GWs \cite{Goldberg1957}. In 1974, Russell Alan Hulse and Joseph Hooton Taylor, Jr. discovered the first binary pulsar, such discovery earned the Nobel Prize in 1993\cite{1609.09400,pulsar1979,pulsar1982}. Pulsar timing observations showed a gradual decay of the orbital period of the Hulse-Taylor pulsar that matched the angular momentum and energy loss in gravitational radiation predicted by general relativity.

Until 2016, an hundred years later after Einstein conjectured GW, Virgo and LIGO Scientific collaboration announced two GW events named GW150914 and GW151226 from two merging black holes (BHs) binary systems \cite{1602.03837,1606.04855}. Furthermore, the third and fourth detected GWs signal, called GW170104 \cite{1706.01812} and GW170814 \cite{1709.09660}, produced by the coalescence of a pair of stellar-mass BHs . Where the GW170814 are the first signal be detected by three detectors (one for Virgo, others are in LIGO) , thus people can located the regime of the signals more precisely. Moreover, the fifth detected GWs signal, GW170817, is the first ever detection of GWs originating from the coalescence of a binary neutron stars system. In contrast to the case of binary BH mergers, binary neutron star mergers were expected to yield an electromagnetic counterpart, that is, a light signal associated with the event. A gamma-ray burst (GRB 170817A) \cite{1803.02768} was detected by the Fermi Gamma-ray Space Telescope, occurring 1.7 seconds after the gravitational wave transient. The signal, originating near the galaxy NGC 4993, was associated with the neutron star mergers. This was corroborated by the electromagnetic follow-up of the event (AT 2017gfo) \cite{1710.05854}, involving 70 telescopes and observatories and yielding observations over a large region of the electromagnetic spectrum which further confirmed the neutron star nature of the merged objects and the associated kilonova. Gracefully, the Nobel Prize in Physics was awarded to Rainer Weiss, Kip Thorne and Barry Barish for their role in the detection of GWs in 2017. In addition, it has been proven by Bondi that GW carries energy when passing through the space-time \cite{Bondi1957,Bondi1959}. Because of the cross section of GWs is very small, people expect that GWs still remind the information under their originated. In general, GWs propagate without attenuation throughout the relativistic era, tracing back to the Planck time, unless encounter unusual conditions simultaneously \cite{Vishniac1982}.

The source of GWs could be classified into two categories, one is cosmological and another is astrophysical origin. For the cosmological class, it has long been conjectured that the GW can be produced in the early universe phenomenon \cite{Carr1980,Rees1983}, such as, inflation and reheating epochs \cite{1312.2284}. For these reasons, such kind of GWs are often called primordial GWs and are believed that they will left some imprint on the CMB. Moreover, during the evolution of our universe, it is expected that various phase transitions (PTs) occurred as the universe cooling down, for instance, QCD PT \cite{1703.02801}, electroweak PT \cite{9310044}, etc. Measurable physical quantities are originated from those symmetric breaking along the PTs. During different PTs, GWs are also expected to be produced with different characteristic features. More generally, the topological defects, i.e., cosmic string \cite{1606.05585}, domain wall \cite{9401007}, etc, will create GWs as well. Therefore, the detection of GWs due to the cosmological origins serve the probing job to reveal modern physics associated with the early universe and also the history we passed by. On the astrophysics class, GWs can be produced in extremely diverse processes, for example, rotation of non-symmetric neutron star, explosion of supernova, merging binary systems, etc. There is a short review \cite{1703.00187}. The corresponding GWs detectors and sources are labeled in Fig.\ref{GW_sources}.
\begin{figure}[t]
\begin{center}
\includegraphics[
height=3.4in, width=4.8in]
{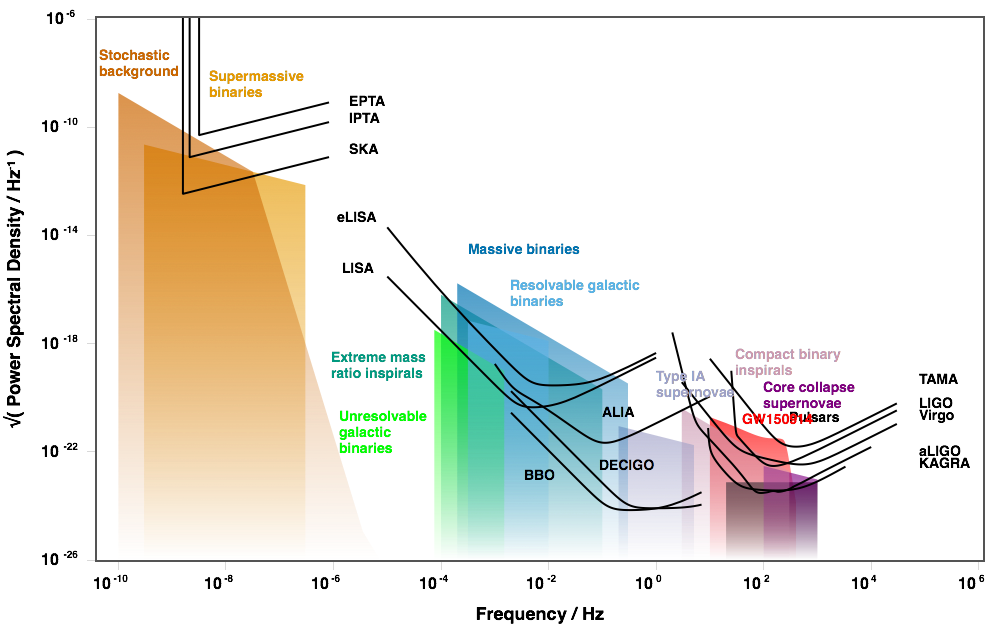}
\end{center}
\caption{The various GW sources and the noise curves of the proposed GW detectors. \cite{1408.0740}.}
\end{figure} \label{GW_sources}

\begin{figure}[t]
\begin{center}
\includegraphics[
height=2.05in, width=2.3in]
{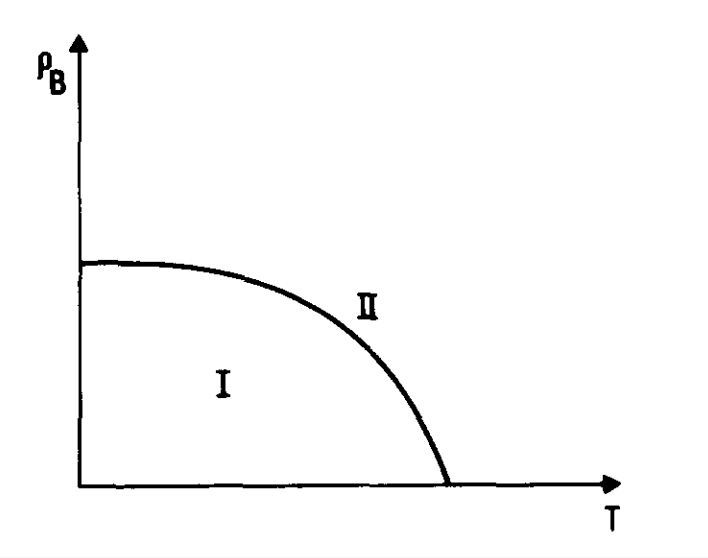}
\includegraphics[
height=2.1in, width=2.3in]
{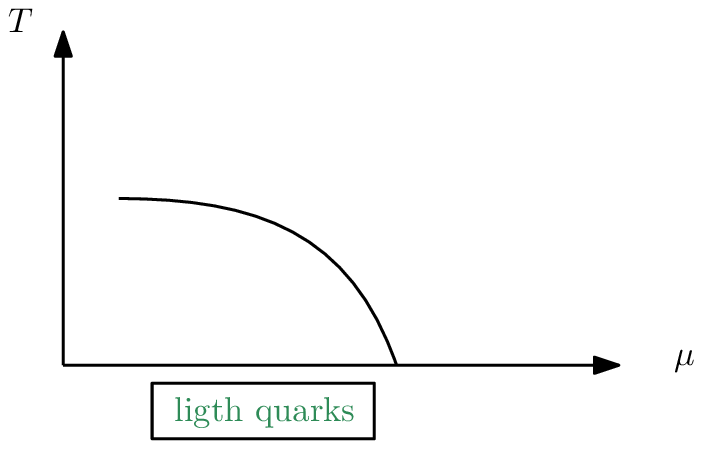}
\includegraphics[
height=2.1in, width=2.3in]
{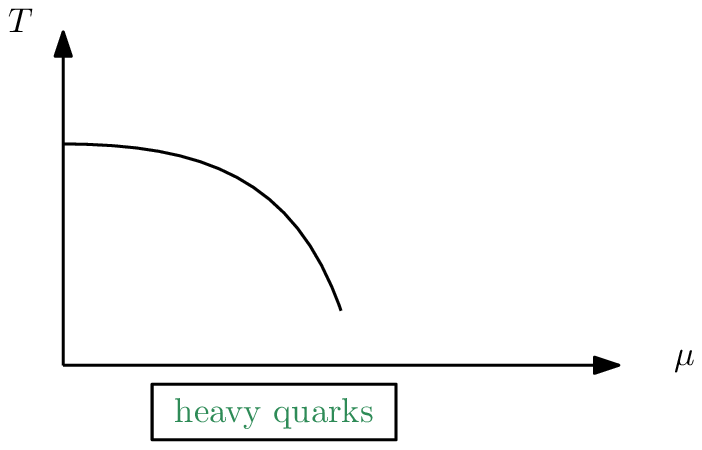}
\vskip -0.05cm \hskip 0.15 cm (a) \hskip 5.5 cm (b) \hskip 5.5 cm (c)
\end{center}
\caption{(a)The schematic QCQ phase diagram of hadronic matter was claimed by Cabibbo and Parisi in 1975, where $\rho_B$ labels the baryonic density. The transition line divided the phase space into two separated phases. Quarks are confined to form a bounded state in phase I and unconfined in phase II \cite{Cabibbo1975}. (b) Nowadays, lattice simulations \cite{1009.4089,1111.4953} convince people that the QCD phase diagram behaves as (a) in chiral limit, but exists a crossover regime for massive cases. Light quarks PT should display as (b), otherwise would present as (c) when considering heavy quarks even the pure gauge limit.} \label{fig_cabibbo1975}
\end{figure}

In order to investigate the GWs through QCD PT, we need to understand the QCD phase diagram as better as possible. It is widely believed that, the quarks are confined to hadronic bound states, e.g. mesons and baryons, at low temperature T and small chemical potential $\mu$ (low quark density) region, while free quarks can exist at high temperature and large chemical potential (high quark density) region. Therefore, it is natural to conjecture that there is a phase transition between the two phases as showed in Fig.\ref{fig_cabibbo1975} (carton phase diagram at m=0). Since QCD itself is a strongly coupled system at low energy, therefore non-perturbative approaches are needed to attack the problem. Many effective theories have been developed to study the low energy QCD, such as chiral lagrangain, Nambu-Jona-Lasinio model, quark-meson model, QCD sum rules, renormalization group, etc., but with different disadvantages. On the other hand, lattice QCD simulation is the currently most promising treatment to solve the strongly coupled system. However, lattice techniques can only well-organized in zero chemical potential case and always encounter sign problem as concerning finite chemical potential. Recently, lattice QCD has developed a few treatments to deal with the finite density situations, for instance, reweighting or imaginary chemical potential method, even $\mu/T$ expansion till high order. Nevertheless, these methods are only reliable on small chemical potential circumstance and are out of control in large chemical potential.
Holographic QCD from gauge/gravity correspondence delivers a new aspect to solve the puzzle of QCD phase structure. AdS/QCD duality is a low energy limit effective theory from supergravity theory which claims that a strongly coupled field theory living on boundary spacetime corresponds to a weakly coupled gravitational theory belongs to the one dimensional higher bulk manifold. In general classification, there are two classes of AdS/QCD models, one is called top-down approaches, which are established by the ingredients of D-branes from supergravity, another is called bottom-up approaches, which are mainly made by phenomenology requirements in model building viewpoint.

Einstein-Maxwell-scalar system is a well-studied bottom-up model. It is useful to study QCD field theory in 5 dimensional gravitational bulk \cite{1301.0385,1406.1865, 1506.05930} or even arbitrary higher dimensional structure inside the geometry \cite{1705.07587}. In this work, we construct a bottom-up holographic QCD soft-wall model by studying its duality of 5-dimensional gravitational theory coupled to a Abelian gauge field and a neutral scalar field. We analytically solve the equations of motion to obtain a family of BH backgrounds which only depend on two arbitrary functions $f(z)$ and $A(z)$. Without loss of generality, in order to well establish a proper gravitational theory, we wonder the energy conditions, such as null energy condition (NEC). In the gravity side, NEC gives a constraint of gauge kinetic function $f(z)$ and warped function $A(z)$. The NEC even support the reality of neutral scalar field which plays the crucial role in the soft-wall construction. One of the advantage properties for the soft-wall holographic QCD models is the vector meson spectrum satisfies the linear Regge trajectories at zero temperature and zero density, we are able to fix the function $f(z)$, which has to satisfied the NEC, by requiring the linear meson spectrum. Then, by choosing a suitable function $A(z)$, we obtain a family of analytically solution of the blackening background which appropriately describe many important properties in QCD. We explore the phase structure of the BH background by studying its thermodynamics quantities under different temperature and chemical potential. The phase diagrams for light and heavy quarks are showed in Fig.\ref{fig_cabibbo1975}.

Cosmological first-order PT could deliver far-reaching consequence, most of them are constructed by the fact that the first-order PT proceeds by bubble nucleation and large portion of the available vacuum energy is stored closed to the bubble walls. The first-order PT is one of the source causing GWs. We use the QCD confinement PT, not only light mesons but heavy mesons as well, to investigate the GWs spectrum. First of all, we consider the runaway situation, which means the bubbles wall velocity approaches speed of light. Then study the no-runaway cases, including detonations and defragrations. In \cite{1705.07587}, the authors proved that the speed of sound in (3+1)-dimensional QCD is always less than the conformal limit in any finite arbitrary chemical potentials. As the last part, we concern the duration of the PT.

The paper is organized as follows. In section \ref{EMS}, we introduce a 5-dim holographic QCD model, called Einstein-Maxwell-scalar (EMS) model, which has been widely investigate to study various PTs and receive productive achievements. Afterwards, we explore the confinement-deconfinement PT in section \ref{BH PTs}. Inserting QCD PT into GWs analysis is arranged in section \ref{GWs}. We conclude our results in the final section.

\section{Deconfinement Phase Transition in Holographic QCD} \label{EMS}
The EMS system has been widely studied in constructing bottom-up holographic QCD models. In this section, we briefly review the EMS system. By solving the equations of motion analytically, we obtain a family of BH solutions which have been extended studied in \cite{1301.0385,1406.1865,1506.05930}. In this paper, we consider two special cases among the family of the solutions which correspond to the holographic QCD models for the light and heavy quarks, respectively.

\subsection{Review of EMS System}
We consider a 5-dimensional EMS system with probe matter fields. The system can be described by an action with two parts, background sector $S_{B}$ and matter sector $S_{m}$. In Einstein frame, the action can be written as,
\begin{eqnarray}
S&=&S_{B}+S_{m}, \label{eq_S}\\
S_{B} &=& \frac{1}{16\pi G_{5}} \int d^{5}x\sqrt{-g}
	\left[{R-\frac{f\left(\phi\right)}{4}F^{2}}
			-\frac{1}{2}\partial_{\mu}\phi \partial^{\mu}\phi
			-V\left(\phi\right) \right], \label{eq_SB_Ef} \\
S_{m} &=& -\frac{1}{16\pi G_{5}}\int d^{5}x\sqrt{-g}
		\left[{\frac{f\left(\phi\right)}{4}}
		\left(F_{V}^{2}+F_{\tilde{V}}^2\right)\right]. \label{eq_Sm_Ef}
\end{eqnarray}where ${G}_{5}$ is the 5-dimensional Newtonian constant, ${{F}}_{\mu\nu} = \partial_{\mu}{A}_{\nu}-\partial_{\nu}{A}_{\mu}$ is the gauge field strength corresponding to the Maxwell field, $f(\phi)$ is the gauge kinetic function associated to Maxwell field and $V(\phi)$ is the potential of the scalar field.
The matter part of the EMD system $S_{m}$ includes two massless gauge fields $A_{\mu}^{V}$ and $A_{\mu}^{\tilde{V}}$, which are treated as probes in the background, describing the degrees of freedom of vector mesons and pseudovector mesons on the 4-dimensional boundary.

Now we are in the stage to derive the equations of motion of our EMS system from Eqs.(\ref{eq_SB_Ef}-\ref{eq_Sm_Ef}). We first study the background by turning off the probe matters of vector field $A_{\mu}^{V}$ and pseudovector field $A_{\mu}^{\tilde{V}}$, the equations of motion can be derived as,
\begin{eqnarray}
\nabla^{2}\phi &=& \frac{\partial V}{\partial\phi}+\frac{F^2}{4}\frac{\partial f}{\partial\phi}, \label{eq_eom_phi}\\
\nabla_{\mu}\left[ f(\phi)F^{\mu\nu} \right] &=&0, \label{eq_eom_A}\\
R_{\mu\nu}-\frac{1}{2} g_{\mu\nu}R &=& \frac{f(\phi)}{2} \left( F_{\mu\rho}F_{\nu}^{~\rho}-\frac{1}{4}g_{\mu\nu}F^{2}\right) +\frac{1}{2}\left[\partial_{\mu}\phi\partial_{\nu}\phi-\frac{1}{2}g_{\mu\nu}\left(\partial\phi\right)^{2}-g_{\mu\nu}V(\phi)\right] \label{eq_eom_g}.
\end{eqnarray}
We consider the following BH ansatz of the background metric in Einstein frame as
\begin{eqnarray}
ds^{2} &=& \frac{e^{2A\left(z\right)}}{z^{2}}
			\left[-g(z)dt^{2} +d\vec{x}^{2} +\frac{dz^{2}}{g(z)}\right],\label{eq_metric}\\
\phi &=& \phi\left(z\right),~ A_{\mu}=\left(A_{t}\left(z\right),\vec{0},0\right), \label{eq_ansatz}
\end{eqnarray}
where $z = 0$ corresponds to the conformal boundary of the 5-dimensional space-time and $g(z)$ stands for the blackening factor. Here we have set the radial of $AdS_5$ to be unit by scale invariant.

The equations of motion Eqs.(\ref{eq_eom_phi}-\ref{eq_eom_g}) leads to the following equations for the background fields,
\begin{eqnarray}
\phi^{\prime\prime}+\left(\frac{g^{\prime}}{g}+3A^{\prime}-\dfrac{3}{z}\right) \phi^{\prime}+\left( \frac{z^{2}e^{-2A}A_{t}^{\prime2}f_{\phi}}{2g}-\frac{e^{2A}V_{\phi}}{z^{2}g}\right)   &=&0,\label{eom_phi}\\
A_{t}^{\prime\prime}+\left(  \frac{f^{\prime}}{f}+A^{\prime}-\dfrac{1}{z}\right)  A_{t}^{\prime}  &=&0,\label{eom_At}\\
A^{\prime\prime}-A^{\prime2}+\dfrac{2}{z}A^{\prime}+\dfrac{\phi^{\prime2}}{6}&=& 0,\label{eom_A}\\
g^{\prime\prime}+ 3g'\left(  A^{\prime}-\dfrac{1}{z}\right) - \frac{fz^2 A_t'^2}{e^{2A}} &=& 0,\label{eom_g}\\
A''+3A'^2-\frac{2}{z}A'+\left( A'-\frac{1}{z} \right) \left( \frac{3g'}{2g}-\dfrac{4}{z}\right) +\dfrac{g''}{6g}+\frac{e^{2A}V}{3z^{2}g}  &=& 0.
\label{eom_V}
\end{eqnarray}
As a theory of gravity, considering energy conditions can make the theory more complete and self-consistent. The null energy condition (NEC) is written as $NEC= T_{\mu \nu}N^{\mu}N^{\nu} \geq 0$, where the null vector is defined as $g_{\mu\nu}N^{\mu}N^{\nu}=0$ and energy momentum tensor is expressed in Eq.(\ref{eq_eom_g}). By choosing the independent null vectors as
\begin{equation}
   N^{\mu}=\frac{1}{\sqrt{g\left(  z\right)  }}N^t
          +\frac{\sin\theta}{\sqrt{3}} N^{\vec{x}}
          + \cos\theta\sqrt{g\left(z\right)} N^{z},
\end{equation}
where $\theta$ is an arbitrary parameter.
Thus the NEC requires
\begin{equation}
    T_{\mu\nu}N^{\mu}N^{\nu}=\frac{1}{2}\left[6g \left(A^{\prime\prime}-A^{\prime2}+\dfrac{2}{z}A^{\prime}\right ) \cos^2\theta+\left(g^{\prime\prime}+ 3g'A'-\dfrac{3g' }{z}\right)\sin^2\theta\right] \geq 0, \label{NEC}
\end{equation}
which, imposed with the above equations of motion, turns out that the gauge kinematic function is non-negative and ensures the reality of the scalar field $\phi$, i.e. $\phi'^2 \geq 0$ and $f \geq 0$.

A set of proper boundary condition is essential role to solve the above equations of motion. In the holographic point of view, boundary conditions are given at UV and IR which is located at boundary of the bulk and the BH horizon in the bulk, respectively. We impose the boundary conditions that the metric in the string frame to be asymptotic to $AdS_5$ at the boundary $z=0$ and the BH solutions are regular at the horizon $z=z_H$.
\begin{enumerate}[(i)]
    \item $z \to 0:$
		\begin{equation}
 		A(0)+\sqrt{\frac{1}{6}}\phi(0)=0,~ g(0)=1. \label{bdy_0}
		\end{equation}
    \item $z=z_H:$
        \begin{equation}
 		A_t(z_H)=g(z_H)=0\footnote{Because of $A_\mu A^\mu=g^{tt}A_t A_t$. Thus, it is a natural physical requirement to suppress $A_{\mu}(z_H) A^{\mu}(z_H)$ converge, since $g^{tt}(z_H)$ is divergent.}. \label{bdy_zh}
		\end{equation}
\end{enumerate}
The equations of motion Eqs.(\ref{eom_phi}-\ref{eom_g}) can be solved analytically as,
\begin{eqnarray}
\phi\left(  z\right) &=&\int_0^z dy\sqrt{-6\left( A^{\prime\prime}
-A^{\prime2}+\dfrac{2}{z}A^{\prime}\right)  },\label{phip-A}\\
A_{t}\left(  z\right)   &=& \mu \left(1-\frac{\int_0^{z}\frac{y}{fe^{A}}dy}{\int_0^{z_H}\frac{y}{fe^{A}}dy}\right)=\mu-\rho z^2 +\cdots,\label{At-A}\\
g\left(  {z}\right)   &  =&1-\frac{\int_{0}^{z} \frac{y^3}{e^{3A}} dy}{\int_{0}^{z_{H}%
} \frac{y^3}{e^{3A}} dy}+\dfrac{\mu^{2}\left\vert
\begin{array}
[c]{cc}%
\int_{0}^{z_{H}} \frac{y^3}{e^{3A}} dy & \int_{0}^{z_{H}}\frac{y^3}{e^{3A}}dy\int_{0}^{y}%
\frac{x}{fe^{A}}dx\\
\int_{z_{H}}^{z}\frac{y^3}{e^{3A}}dy & \int_{z_{H}}^{z}\frac{y^3}{e^{3A}}dy\int_{0}^{y}%
\frac{x}{fe^{A}}dx
\end{array}
\right\vert }{ \left( \int_{0}^{z_{H}} \frac{y^3}{e^{3A}} dz \right) \left( \int_{0}^{z_{H}} \frac{z}{fe^{A}}dz \right)^{2}}, \\
V(z) &=&-\frac{3g z^2}{e^{2A}}
        \left[
            \left( A''+3A'^2-\frac{2}{z}A' \right)
            +\left( A'-\frac{1}{z} \right) \left( \frac{3g'}{2g}-\dfrac{4}{z}\right) +\dfrac{g''}{6g}
        \right]. \label{V-A}
\end{eqnarray}
where $\mu$ is the baryon chemical potential, by the holographic dictionary of the gauge/gravity correspondence, and $\rho$ is the baryon density that relates to the chemical potential as,
\begin{equation}
\rho=\dfrac{\mu}{2 \int_{0}^{z_{H}}\dfrac{y}{fe^A}dy}. \label{rho}%
\end{equation}
In the solution (\ref{phip-A}-\ref{V-A}), the different choice of the functions
$A\left(  z\right)  $ and $f\left(  z\right)  $ corresponds to the different BH solutions. However, NEC constrains the choice of the functions $A\left(  z\right)  $ and $f\left(  z\right)  $ in Eq.(\ref{NEC}).

\subsection{Meson Mass Spectrum}
As mentioned in the introduction, one of the crucial properties for the soft-wall holographic QCD models is that the vector meson spectrum satisfies the linear Regge trajectories at zero temperature and zero density, i.e. $\mu=0$. This issue was first addressed in the soft-wall model \cite{0602229} using the method of AdS/QCD duality.

We consider the 5-dimensional probe vector field $V$ in the action Eq.(\ref{eq_Sm_Ef}). The equation of motion for the vector field reads
\begin{equation}
\nabla_{\mu}\left[ f\left( \phi\right) F_{V}^{\mu\nu}\right]  ={{0.}}%
\end{equation}
By fixing the gauge $V_{z}=0$, the equation of motion of the transverse part of the vector field $V_{\mu}$ $\left( \partial^{\mu}V_{\mu}=0\right) $ in the background Eq.(\ref{eq_metric}) reduces to%
\begin{equation}
-\psi_{i}^{\prime\prime}+U\left( z\right) \psi_{i}=\left( \dfrac{\omega
^{2}}{g^{2}}-\dfrac{p^{2}}{g}\right) \psi_{i}, \label{eqv}%
\end{equation}
where we have performed the Fourier transformation for the vector field
$V_{i}$ as%
\begin{equation}
V_{i}\left( x,z\right) =\int\dfrac{d^{4}k}{\left( 2\pi\right)^{4}%
}e^{ik\cdot x}v_{i}\left( z\right) , \label{V-v}%
\end{equation}
and
\begin{equation}
v_{i}=\left( \dfrac{z}{e^{A}fg}\right)^{1/2}\psi_{i}\equiv X\psi_{i},
\end{equation}
with the potential function%
\begin{equation}
U\left( z\right) =\dfrac{2X^{\prime2}}{X^{2}}-\dfrac{X^{\prime\prime}}{X}.
\end{equation}
In the case of zero temperature and zero chemical potential, we expect that the discrete spectrum of the vector mesons obeys the linear Regge trajectories. The above Eq.(\ref{eqv}) reduces to a Schr\"{o}dinger
equation%
\begin{equation}
-\psi_{i}^{\prime\prime}+U\left(  z\right)  \psi_{i}=m^{2}\psi_{i},
\label{Schrodinger}%
\end{equation}
where $-m^{2}=k^{2}=-\omega^{2}+p^{2}$. To produce the discrete mass spectrum with the linear Regge trajectories, the potential $U\left( z \right)$ should be in certain forms. A simple choice is to fix the gauge kinetic function as
\begin{equation}
f\left( z\right) = e^{\pm cz^{2}-A\left(  z\right)  },
\end{equation}
which leads the potential to be
\begin{equation}
U\left( z\right) = -\dfrac{3}{4z^{2}}-c^{2}z^{2}. \label{potential}
\end{equation}
The Schr\"{o}dinger Eq.(\ref{Schrodinger}) with the potential in
Eq.(\ref{potential}) has the discrete eigenvalues
\begin{equation}
m_{n}^{2}=4cn, \label{mass}%
\end{equation}
which is linear in the energy level $n$ as we expect for the vector spectrum at zero temperature and zero density which was well known as the linear Regge trajectories \cite{0507246}. It is worth to emphasize that, the mass tower of the considering linear vector meson is only depends on a parameter $c$ in this proper choice of the gauge kinetic function $f(z)$.

For the holographic QCD model of light quarks, by fitting the mass Regge spectrum with $\rho$ meson tower, we fixed the parameter $c=0.227 GeV$. While for the model of heavy quarks, by fitting the mass Regge spectrum with the $j/\psi$ quarkonium states, we fixed the parameter $c=1.16 GeV$.

Once we fix the gauge kinetic function $f(z)$, the equations of motion Eqs.(\ref{eom_At}-\ref{eom_V}) can be analytically solved as
\begin{eqnarray}
\phi^{\prime}\left( z \right) &=&\sqrt{-6\left( A^{\prime\prime}
-A^{\prime2}+\dfrac{2}{z}A^{\prime}\right) }, \\
A_{t}\left(z\right) &=& \mu \left(1-\frac{1-e^{cz^{2}}}{1-e^{cz_{H}^{2}}}\right), \\
g\left(z\right) &=& 1-\dfrac{\int_{0}^{z}\frac{y^{3}}{e^{3A}}dy}{\int_{0}^{z_{H}}\frac{y^{3}}{e^{3A}}dy}
 +\frac{2c\mu^{2}\left\vert
\begin{array}
[c]{cc}%
\int_{0}^{z_{H}}\dfrac{y^{3}}{e^{3A}}dy & \int_{0}^{z_{H}}\dfrac{y^{3}}{e^{3A-cy^2}}dy\\
\int_{z_{H}}^{z}\dfrac{y^{3}}{e^{3A}}dy & \int_{z_{H}}^{z}\dfrac{y^{3}}{e^{3A-cy^2}}dy
\end{array}
\right\vert}{\left(  1-e^{cz_{H}^{2}}\right)^{2} \left( \int_{0}^{z} \frac{y^3}{e^{3A}} dy \right) },\\
V(z) &=&-\frac{3g z^2}{e^{2A}}
        \left[
            \left( A''+3A'^2-\frac{2}{z}A' \right)
            +\left( A'-\frac{1}{z} \right) \left( \frac{3g'}{2g}-\dfrac{4}{z}\right) +\dfrac{g''}{6g}
        \right].
\end{eqnarray}

Eqs.(\ref{phip-A}-\ref{V-A}) represent a family of solutions for the BH background depending on the warped factor $A\left(z\right)$, which could be arbitrary function satisfies the boundary condition in Eq.(\ref{bdy_0}). However, null energy constraint the expression under the squared root in Eq.(\ref{phip-A}) is positive for $z\in(z,z_H)$ to guarantee a real scalar field $\phi(z)$.

For the holographic QCD model of light quarks, we choose the warped factor $A \left( z \right)= -a \ln (bz^2+1)$ with $a=4.046$ and $b=0.01613$. For the model of heavy quarks, we choose the warped factor $A (z)= -cz^2/3-bz^4 $ with $b=0.273 GeV$. The parameters in the warped factors are fixed by comparing the PT temperature with lattice QCD simulations.

\subsection{Phase Structure of the Black Hole Background} \label{BH PTs}
In this section, we review the phase structure of the BH background in Eqs.(\ref{phip-A}-\ref{V-A}) which we obtained in the last section. The entropy density and the Hawking temperature can be calculated as,
\begin{eqnarray}
s&=&\frac{e^{3A(z_H)}}{4z_H^{3}},\\
T&=&\dfrac{z_{H}^{3}e^{-3A\left( z_{H} \right) }} {4\pi \int_{0}^{z_{H}} \frac{y^{3}}{e^{3A}}dy}
    \left[ 1-2c\mu^{2}
    \frac{ e^{cz_{H}^{2}}\int_{0}^{z_{H}}\frac{y^{3}}{e^{3A}}dy-\int_{0}^{z_{H}}\frac{y^{3}}{e^{3A-cy^{2}}}dy  } {\left( 1-e^{cz_{H}^{2}} \right)^{2}}
    \right]  .
\end{eqnarray}
The free energy in grand canonical ensemble can be obtained from the first law of thermodynamics ,
\begin{equation}
F=\epsilon-Ts-\mu\rho, \label{eq_def_F}
\end{equation}
where $\epsilon$ labels the internal energy density. Comparing the free energies between different sizes of BHs at the same temperature under certain finite value of chemical potential, we are able to obtain the phase structure of BHs which corresponds to the phase structure in the holographic QCD due to AdS/CFT correspondence. The free energy in grand canonical ensemble can be calculated from the first law of thermodynamics. At fixed volume, we have
\begin{equation}
dF=-sdT-\rho d\mu.
\end{equation}
Thus, the free energy for a given chemical potential $\mu$ can be evaluated by an integral,
\begin{equation}
F= -\int sdT= \int_{zH}^\infty s(z_H)T'(z_H)dz_H, \label{eq_int_F}
\end{equation}
where we have normalized the free energy of the BHs vanishing at $z_H \to \infty$, i.e. $T=0$, which is equal to the free energy of the thermal gas.

\subsubsection{Phase Transition for Light Quarks}
\begin{figure}[t!]
\begin{center}
\includegraphics[
height=2.6in, width=3.2in]
{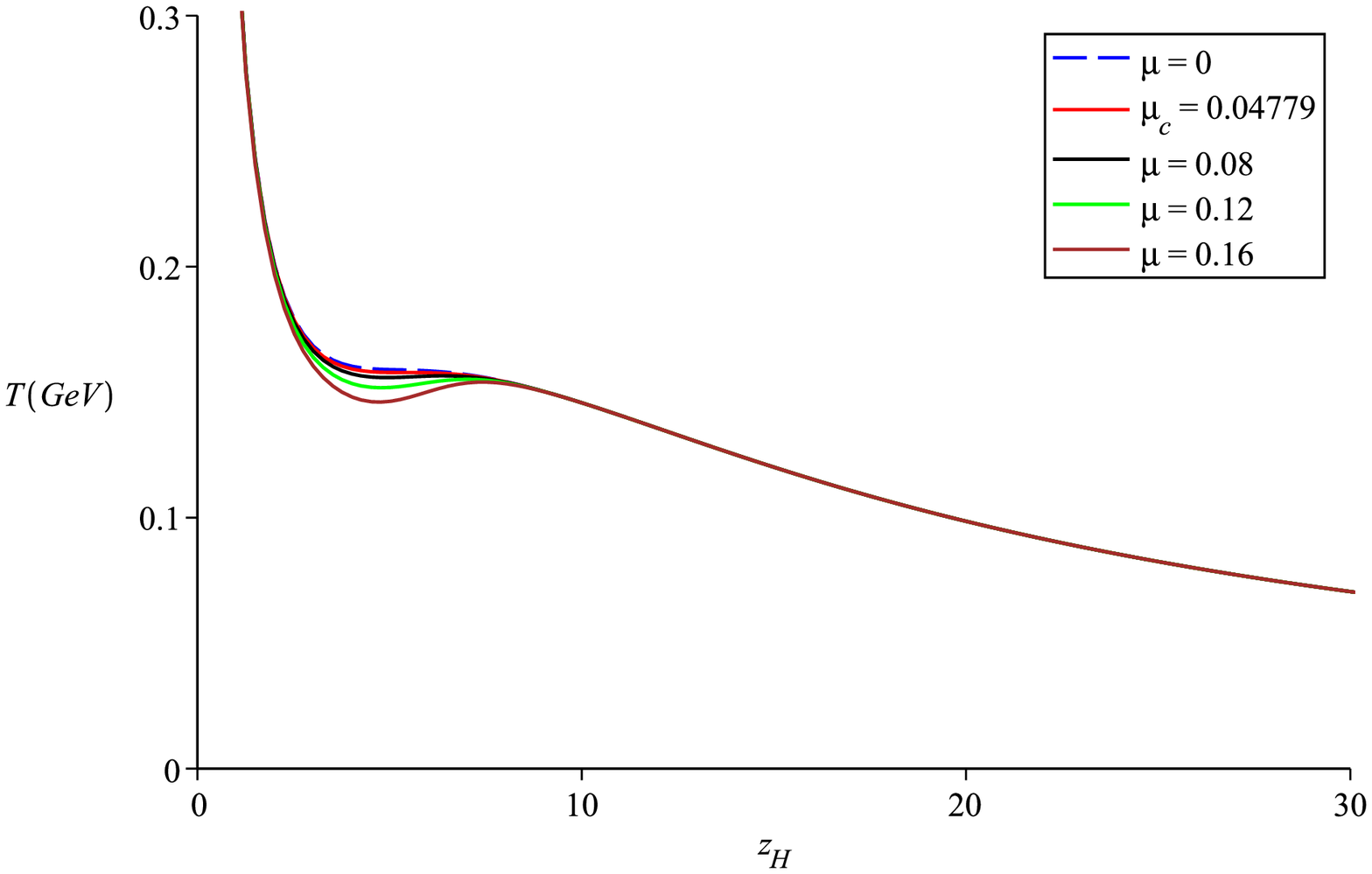}
\includegraphics[
height=2.6in, width=3.2in]
{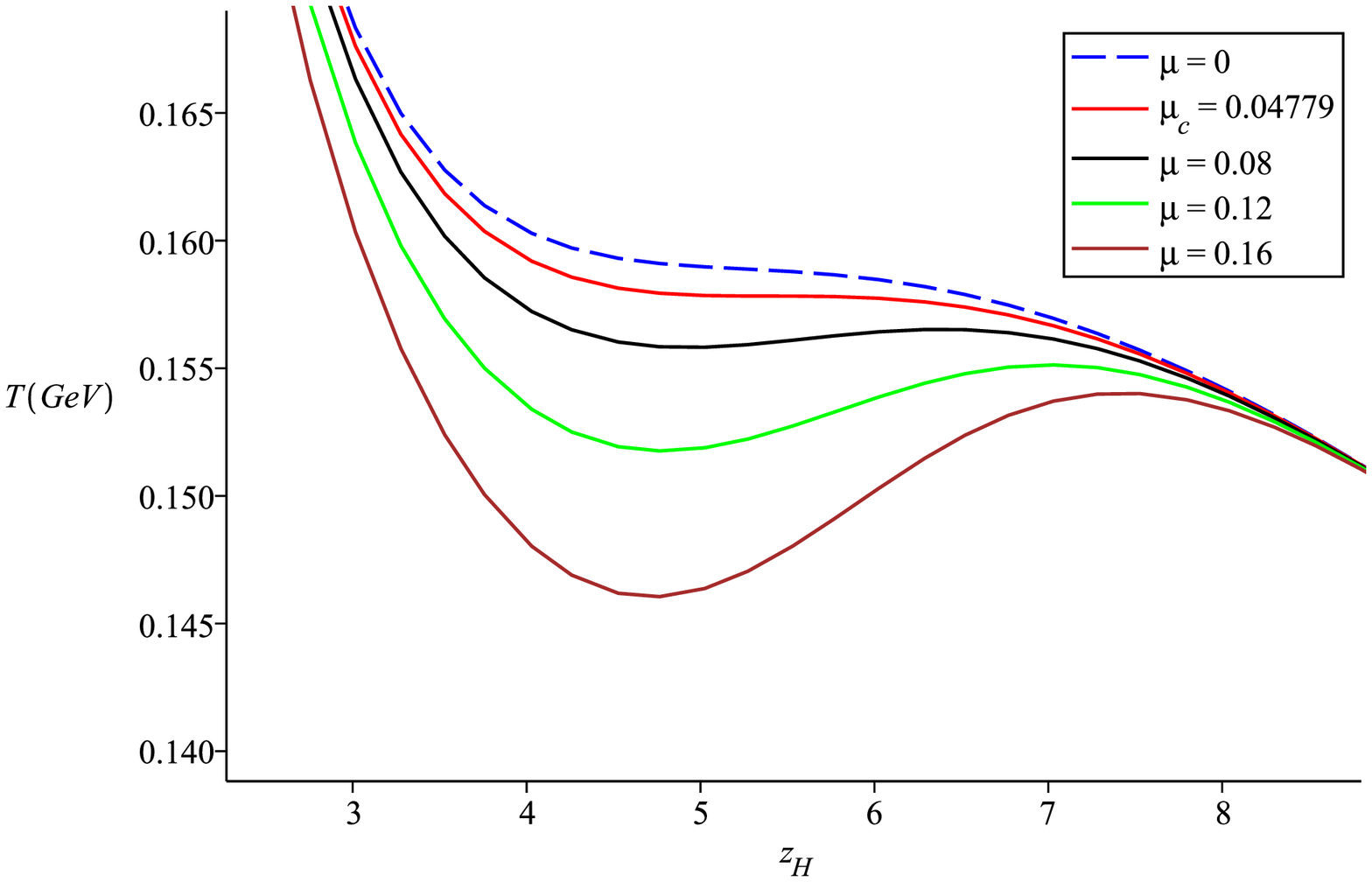}
\vskip -0.05cm \hskip 0.15 cm (a) \hskip 5.5 cm (b)
\end{center}
\caption{(a) The BH temperature v.s. horizon at different chemical potentials, where $\mu_c=0.04779~GeV$. We enlarge the PT region in (b) to display the detail structure.} \label{fig_T}
\end{figure}

The temperature as the function of the horizon, at various chemical potentials, are plotted in Fig.\ref{fig_T}. At small chemical potential, $0 \leq \mu < \mu_c$, the temperature behaves as a monotonous function of horizon and decreases to zero at infinity horizon. Manifestly, there is no PT because of the monotonous behavior of temperature . At large chemical potential, $\mu \geq \mu_c$, the temperature becomes multivalued which implies that there will be a PT between BHs with different sizes.

\begin{figure}[t!]
\begin{center}
\includegraphics[
height=2.2in, width=3.4in]
{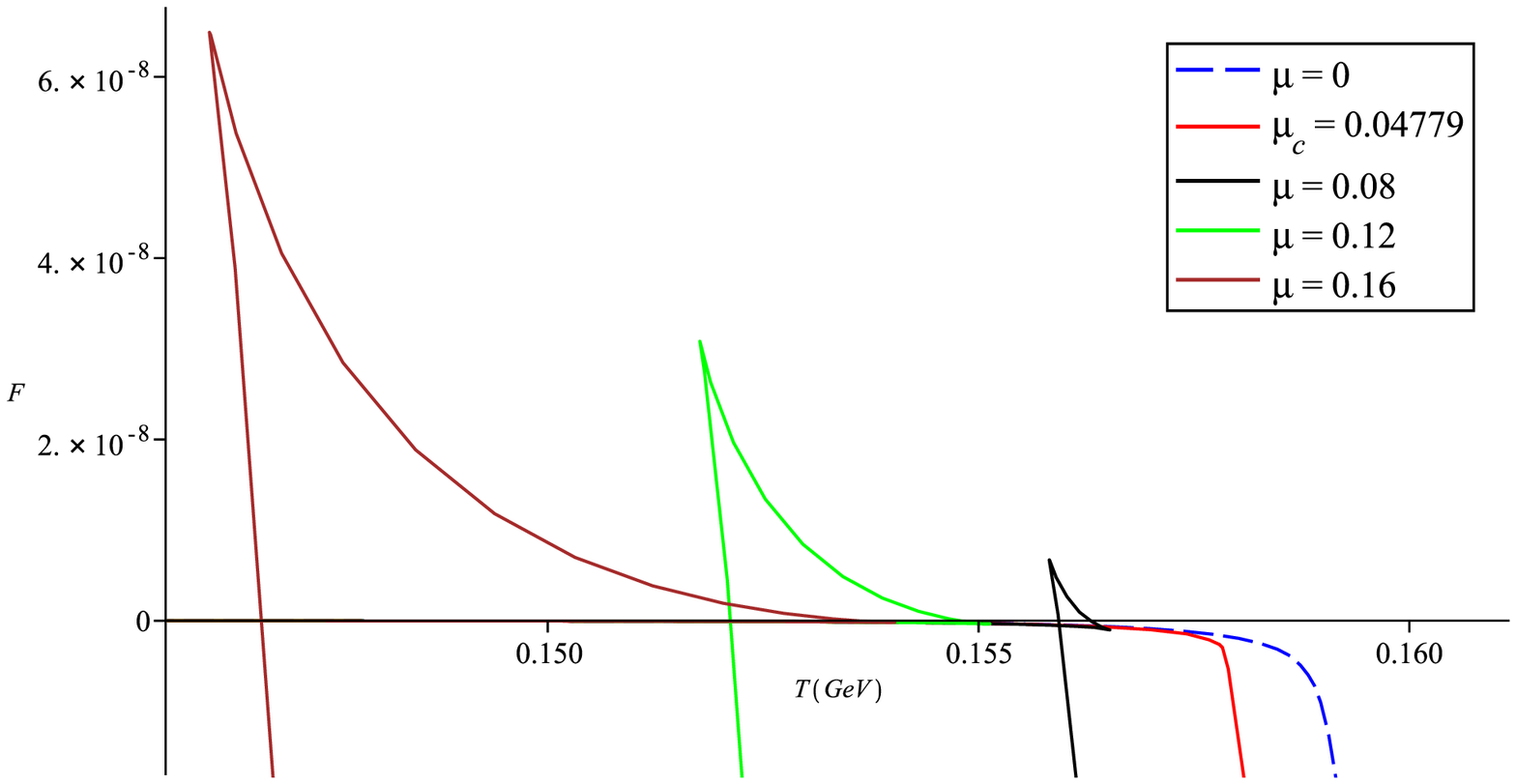}
\includegraphics[
height=2.2in, width=3.4in]
{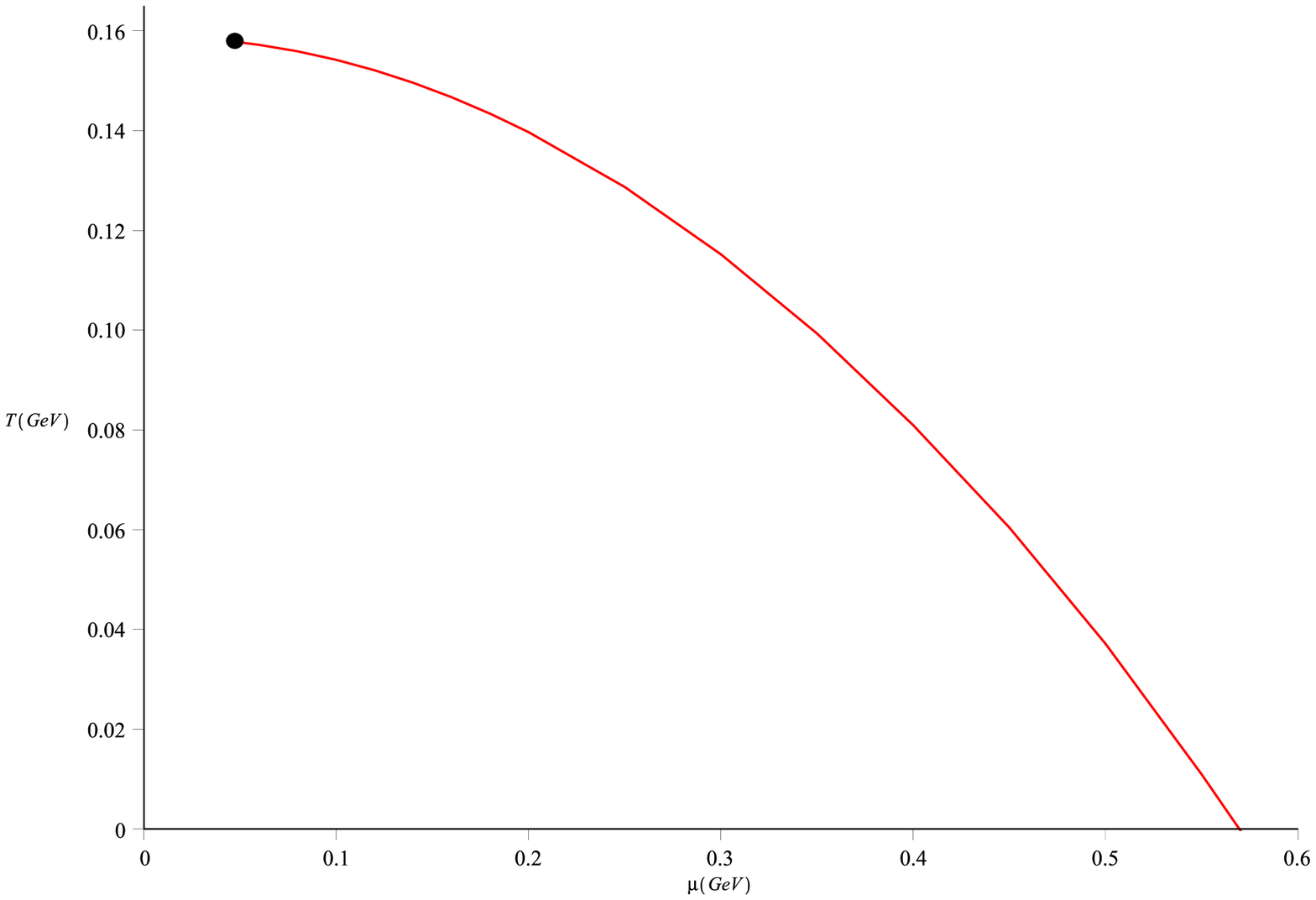}
\vskip -0.05cm \hskip 0.15 cm (a) \hskip 5.5 cm (b)
\end{center}
\caption{(a) The free energy v.s. temperature at various chemical potentials. The free energy behaves as a multivalued function of temperature with the swallow-tailed shapes at $\mu>\mu_c$ and becomes monotonous at $\mu<\mu_c$.
(b) The phase diagram in $T-\mu$ plane. For the large chemical potential $\mu>\mu_c$, the system undergoes a first-order PT at a finite temperature and ends at the critical endpoint $(\mu_c,T_c) \simeq (0.04779,0.1578)$. For small chemical potential $0\leq \mu < \mu_c$, the PT reduceds to a crossover. The zero temperature PT is located at $\mu_{T=0}=0.5695$.} \label{fig_FT} \label{fig_Tmu_BH}
\end{figure}

The free energy v.s. temperature at various chemical potentials is plotted in Fig.\ref{fig_FT}(a). The intersection of free energy implies that there exists a PT between two BHs with different sizes at the PT temperature. For $\mu>\mu_c$, the free energy behaves as the swallow-tiled shape and shrinks into a singular point at $\mu=\mu_c$, then disappears for $\mu<\mu_c$. The behavior of the free energy implies that the system undergoes a first-order PT at each fixed chemical potential $\mu>\mu_c$ and ends at the critical endpoint at $(\mu_c,T_c)$ where the PT becomes second order. For $\mu<\mu_c$, the PT reduces to a crossover.

The phase diagram of BH to BH PT is plotted in Fig.\ref{fig_Tmu_BH}(b).

\subsubsection{Phase Transition for Heavy Quarks}
\begin{figure}[t]
\begin{center}
\includegraphics[
height=2.2in, width=3.2in]
{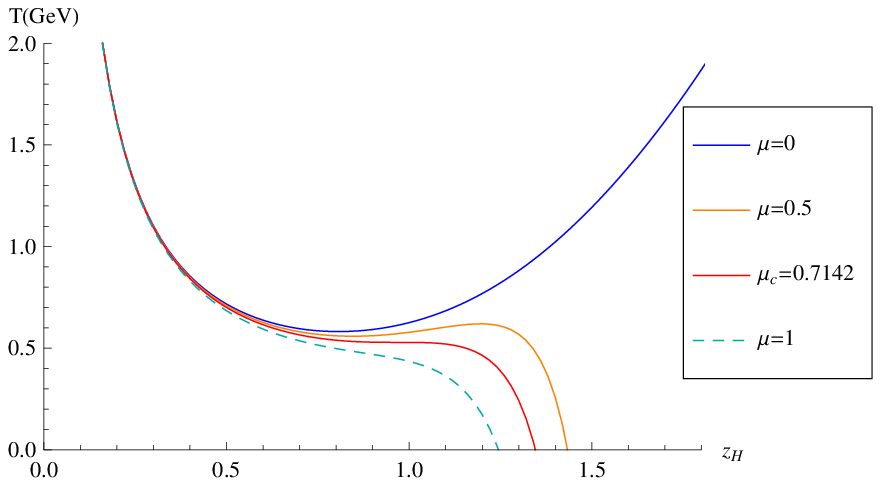}
\includegraphics[
height=2.2in, width=3.2in]
{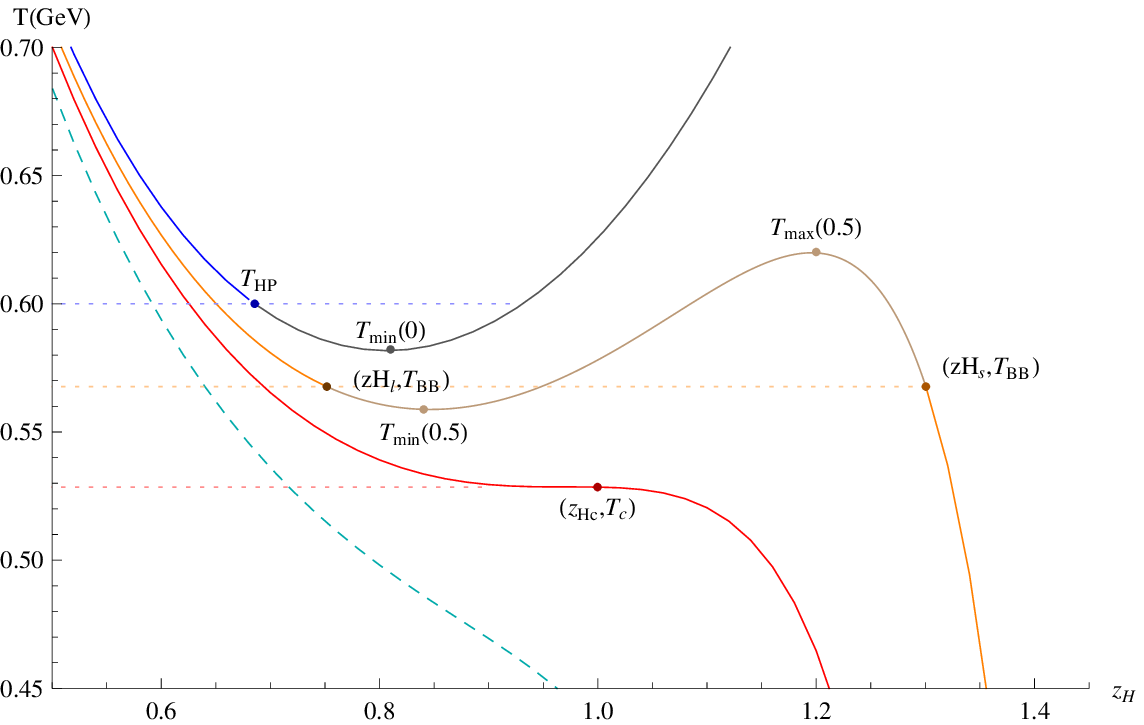}
\vskip -0.05cm \hskip 0.15 cm (a) \hskip 5.5 cm (b)
\end{center}
\caption{(a) The BH temperature v.s. horizon at different chemical potentials, where $\mu_c=0.714~GeV$. We enlarge the PT region in (b) to display the detail structure.} \label{fig_heavyT}
\end{figure}
\begin{figure}[t]
\begin{center}
\includegraphics[
height=2.4in, width=3.4in]
{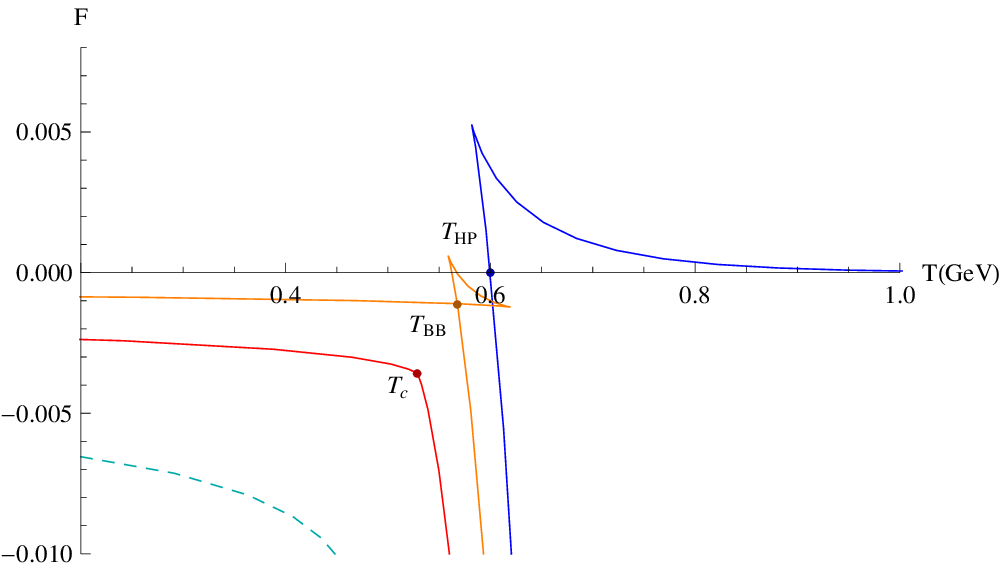}
\includegraphics[
height=2.4in, width=3.4in]
{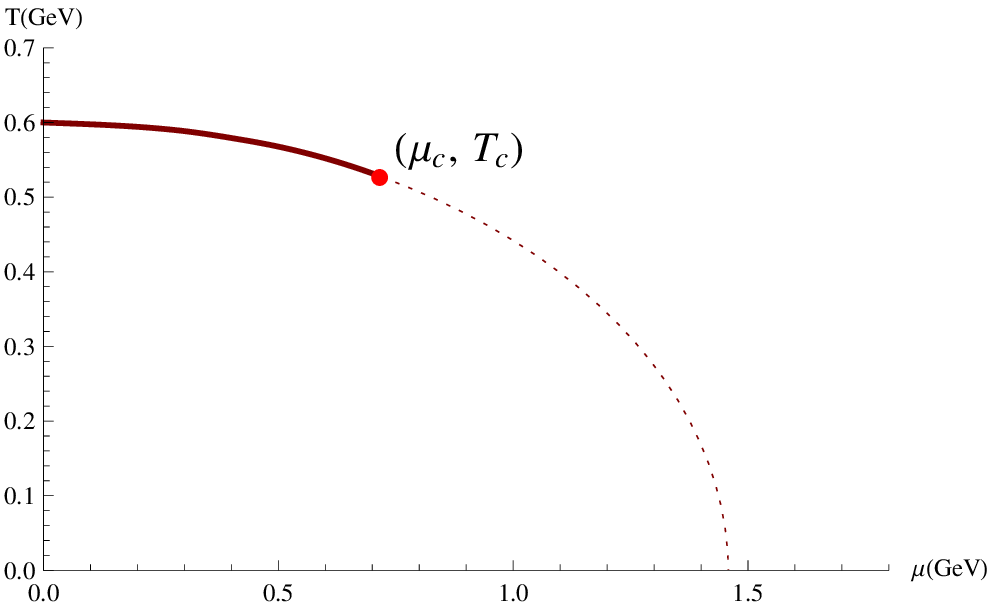}
\vskip -0.05cm \hskip 0.15 cm (a) \hskip 5.5 cm (b)
\end{center}
\caption{(a) The free energy v.s. temperature at chemical potentials $\mu=0,0.5,0.714,1 GeV$. At $\mu=0$, the free energy intersect with the $x$-axis at $T=T_{HP}$ where the BH dissolves to thermal gas by Hawking-Page PT. For $0<\mu<\mu_{c}\simeq0.714GeV$, the temperature reaches its maximum value where the free energy turns back to intersect with itself at $T=T_{BB}$ where the BH to BH transition happens. For $\mu>\mu_{c}$, the swallow-tailed shape disappears and there is no PT in the BH background. (b) The phase diagram in $T$ and $\mu$ plane. At small $\mu$, the system undergoes a first-order PT at finite $T$. The first-order PT stops at the critical point $(\mu_{c},T_{c})=(0.714GeV,0.528GeV)$, where the PT becomes second-order. For $\mu>\mu_{c}$, the system weaken to a sharp but smooth crossover \cite{1301.0385}.} \label{fig_heavyFT} \label{fig_heavyTmu}
\end{figure}
The temperature and free energy are plotted in Fig.\ref{fig_heavyT} and Fig.\ref{fig_heavyFT}. The free energy v.s. temperature at various chemical potentials is plotted in Fig.\ref{fig_heavyFT}(a). For $\mu<\mu_c$, the free energy behaves as the swallow-tiled shape and shrinks into a singular point at $\mu=\mu_c$, then disappears for $\mu>\mu_c$. The behavior of the free energy implies that the system undergoes a first-order PT at each fixed chemical potential $\mu<\mu_c$ and ends at the critical endpoint at $(\mu_c,T_c)$ where the PT becomes second-order. For $\mu>\mu_c$, the PT reduces to a crossover.

The phase diagram of BH to BH PT is plotted in Fig.\ref{fig_Tmu_BH}(b). It is clear that compared to the light quark sector, the phase diagram behaves up-side-down, as we briefly introduce in the previous section and recall Fig.\ref{fig_cabibbo1975}. In the heavy quark scenario, as the temperature cooling down, the first-order PT will become the second-order at the critical end point of the PT, and finally weaken to crossover. The details are explained in our previous works \cite{1301.0385,1506.05930}.

\section{Gravitational Waves} \label{GWs}
In general, GWs are radiated by objects whose motion involves acceleration and its change, provided that the motion is not perfectly spherically symmetric (like an expanding or contracting sphere) or rotationally symmetric (like a spinning disk or sphere). The heavier the objects, and the faster it tumbles, the greater is the gravitational radiation it will give off. More technically, the second time derivative of the quadrupole moment (or the $l_{th}$ time derivative of the $l_{th}$ multipole moment) of an isolated system's stress–energy tensor must be non-zero in order for it to emit gravitational radiation. This is analogous to the changing dipole moment of charge or current that is necessary for the emission of electromagnetic radiation.

The frequency of GWs is characterized by the time scale of its dynamical system, for instance, the frequency of a binary system of two entangled celestial objects correspondences to the frequency they emitted. In principle, gravitational waves could exist at any frequency. Usually, people can classify the GWs in frequency zones. However, very low frequency waves would be impossible to detect and there is no credible source for detectable waves of very high frequency. Stephen Hawking and Werner Israel list different frequency bands for gravitational waves that could plausibly be detected, ranging from $10^{−7} Hz$ up to $10^{11} Hz$ \cite{Hawking1979}. The source of GWs can be recognized from big bang, inflation, gravitational collapses, Gamma-ray burst, supernovae, spinning neutron stars, pulsars, binary systems, extreme mass ratio inspiral, ets.
\begin{center}
\begin{tabular}{ l|l }
 frequency [Hz]               & Sources                                      \\
 \hline \hline
  $10^0 \sim 10^{ 5}$         & binary BHs, binary neutron star, supernovae, first-order PT   \\
  $10^{-6} \sim 10^0$         & supermassive BHs, binary star systems composed of white dwarfs, first-order PT \\
  $10^{-9} \sim 10^{-6}$      & supermassive BHs, pulsar, cosmic string cusp, inflation, first-order PT \\
  $10^{-18} \sim 10^{-15}$    & cosmic microwave background radiation, inflation \\
\end{tabular}
\end{center}

GW from BHs merging has been observed in Laser Interferometer Gravitational-Wave Observatory (LIGO). The frequency of the GW from the binary systems is around $10^0 \sim 10^5 Hz$ , which can be detected by the ground interferometers, such as LIGO and Virgo. On the other hand, the frequency of the primordial GW created by cosmic inflation during the early universe is around $10^{-16} \sim 10^{-18} Hz$, which can be detected by cosmology microwave background polarization. Besides the above two GW sources, it has been argued that the first-order PT during the universe expansion can generate GW as well. The frequency of the GW from PTs is around $10^{-1} \sim 10^{-9} Hz$. For the frequency within $10^{-1} \sim 10^{-5} Hz$, space interferometers, which have longer arms than the ground interferometers, are proposed to accomplish the measurement, such as Laser Interferometer Space Antenna (LISA) and Evolved Laser Interferometer Space Antenna (eLISA). For the frequency within $10^{-6} \sim 10^{-9} Hz$, milisecond pulsar timing array is proposed to detect the GW, such as European Pulsar Timing Array (EPTA), International Pulsar Timing Array (IPTA) and Square Kilometre Array (SKA). In this paper, we focus on the GW from PTs. We will calculate the GW spectrum in various physics conditions and determine whether they can be detected by LISA/eLISA or EPTA/IPTA/SKA.

It has been argued that the first-order PT from a symmetric high temperature phase to a symmetry-breaking low temperature phase will generate GW \cite{1703.00187}. As the temperature decreasing, the unstable false vacuum transfers to the stable true vacuum that creates some vacuum bubbles with the bubble walls.  The vacuum bubbles will expand because of the vacuum energy released from the initial phase. The expanding speed of a bubble wall $v_b$ depends on the friction by the plasma and the difference between the inside pressure and the outside pressure of the bubble wall. As the bubbles expanding, the inside speed $v_-$ and the outside speed $v_+$ are given by \cite{1004.4187}
\begin{equation}
    v_+=\frac{1}{1+\alpha}\left[\left(\frac{v_-}{2}+\frac{1}{6v_-}\right)\pm\sqrt{\left(\frac{v_-}{2}+\frac{1}{6v_-}\right)^2-\frac{1}{3}+\alpha^2+\frac{2}{3}\alpha}\right], \label{vpm}
\end{equation}
where the $\pm$ sign in Eq.(\ref{vpm}) corresponds to the detonation with $v_-<v_+=v_b$ and the deflagration with $v_+<v_-=v_b$ respectively, $\alpha$ labels the strength of the PTs and is one of the most important parameters in identifying the properties of the GW spectrum \cite{1007.1218}. $\alpha$ is defined as the ratio of the vacuum energy density to the thermal energy density,
\begin{align}
    \alpha &=\frac{\epsilon_*}{\frac{\pi^2}{30} g_* T_*^4},\\
    \epsilon_*&=\left[-\Delta F(T) + T \frac{d}{dT}\Delta F(T)\right]_{T=T_*}, \label{eq_epsilon_*}
\end{align}
 $T_*$ is the temperature at which the PT takes place, the number of effective relativistic degree of freedom at the QCD PT is about $g_*\simeq 10$ 
and $\epsilon_*$ refers to the latent heat density at the PT where $\Delta F$ labels the free energy diﬀerence between the two phases.

As we mentioned in Eq.(\ref{vpm}), in the detonation case, the speed of bubble wall is always greater than the speed of sound, while in the deflagration case, the bubble wall speed is always less than the speed of sound. Furthermore, if the friction by the plasma is too small to prevent the wall from accelerating, the bubble wall speed will reaches the speed of light eventually, which corresponds to the case of runaway.

The possibility that the QCD first-order PT produces a detectable GW signal was first proposed by Witten in 1984 \cite{Witten1984}. As we see in Fig.\ref{fig_Tmu_BH}, for $\mu<\mu_c$, the deconfinement transition is a crossover, while for $\mu>\mu_c$, the transition becomes first-order. It has been shown that the detectable GW might be produced as long as the PT is sufficiently strong and lasts for a long time. The GW production can be observed in the pulsar timing experiments \cite{1007.1218}. Latent heat, refers to Eq.(\ref{eq_epsilon_*}) and Eq.(\ref{eq_int_F}), is a indicator which reveals that there is a mount of difference between the energy densities of the two phases in first-order PT. Transition temperature $T_*$ is the temperature where the two phases have the equal pressure and can coexist. As the universe expands, it lost energy, and the universe remains at $T_*$ until the full latent heat of the transition is eliminated.

The GW created from the bubble percolation includes three contributions: bubble collision, sound waves and turbulent Magnetohydrodynamic (MHD). In the following, we will give the numerical formulae to calculate the GW for each contribution.

\subsection{Bubbles Collision}
Bubble collisions was first realized by Witten in 1984 \cite{Witten1984}. It can be studied by using the envelope approximation \cite{0806.1828}. The numerical fits gives the energy density of the GW from bubbles collision,
\begin{equation}
    h^2\Omega_{en}=3.5 \times 10^{-5}\left(\frac{H_{\ast}}{\tau}\right)^2\left(\frac{\kappa\alpha}{1+\alpha}\right)^2\left(\frac{10}{g_{\ast}}\right)^{\rfrac{1}{3}} \left(\frac{0.11v^{3}_{b}}{0.42+v^{2}_{b}}\right) S_{en}(f),
\end{equation}
where the spectral shapes of the GW by the numerical fits is
\begin{equation}
    S_{en}(f)=\frac{3.8(\frac{f}{f_{en}})^{2.8}}{1+2.8(\frac{f}{f_{en}})^{3.8}},
\end{equation}
which is almost a function of $f^3$ for small frequencies and $f^{-1}$ for frequencies larger than the peak frequency. The present red-shifted peak frequency $ f_{en}$ is expressed as
\begin{align}
    f_{en}(f) &= 1.13 \times10^{-8} \left(\frac{f_*}{\tau}\right) \left(\frac{\tau}{H_*}\right) \left(\frac{T_*}{100}\right) \bigg( \frac{g_*}{10} \bigg) ^{\rfrac{1}{6}}, \\
    \frac{f_{\ast}}{\tau} &=\frac{0.62}{1.8-0.1v_b+v^2_b},\\
    H_* &=\sqrt{\frac{8\pi^3g_*}{90}}\frac{T_*^2}{m_{pl}},
\end{align}
where $\tau^{-1}$ is the duration of the PT, $H_*$ is Hubble parameter at the PT and the Planck mass $m_{pl}=1.22\times10^{22}$ MeV. In this paper, we mainly use $\tau=H_*$ to demonstrate our results. However, we still remains for tuning different proportion of $H_*$ in our following discussion in Fig.\ref{fig_difftau}.
Bubbles produced by quantum tunneling grow with wall the velocity $v_b$. It is possible for $v_b$ approaches to the speed of light, which called runaway configuration. The criterion for runaway bubbles is the value of $\alpha$ compared with $\alpha_\infty$ (the minimum value of α for runaway bubbles),
\begin{equation}
    \alpha_{\infty}=\frac{30}{24\pi^2}\frac{\sum_{a} c_{a} \Delta m^2_{a}}{g_* T^2_{*}},
\end{equation}
where $c_{a}=1 \left( 1/2 \right) N_{a}$ is the degrees of freedom for bosonic (fermionic) species and $\Delta m_a$ is the mass difference of the particle between two phases. We choose $\Delta m_a \simeq 300$ MeV and $N_{a}=6$ as in \cite{1011.1749}.
For relativistic velocities, the efficiency factor $\kappa$ , that measures the fraction of the vacuum energy converted to the kinetic energy of the bubbles, is defined as \cite{1004.4187}
\begin{equation}
    \kappa=1-\frac{\alpha_{\infty}}{\alpha}.
\end{equation}
For non-relativistic velocities,  the GW from bubbles collision can be neglected.

\subsection{Sound Waves and MHD Turbulence}
Apart from bubbles collision, there are two processes involved in production of the GW, sound waves and MHD turbulence. Hence, the contribution of the GW energy density from sound waves and MHD turbulence are
\begin{align}
    &h^2\Omega_{sw}=5.7 \times 10^{-6} \left( \frac{H_{\ast}}{\tau} \right) \left( \frac{\kappa_{v}\alpha}{1+\alpha} \right)^2 \left( \frac{10}{g_{\ast}}\right )^{\rfrac{1}{3}} v_b S_{sw}(f),\\
    &h^2\Omega_{tu}=7.2 \times 10^{-4} \left(\frac{H_{\ast}}{\tau}\right) \left(\frac{\kappa_{tu}\alpha}{1+\alpha}\right)^{\rfrac{3}{2}} \left(\frac{10}{g_{\ast}}\right)^{\rfrac{1}{3}} v_b S_{tu}(f).
\end{align}
The spectral shapes of the GW from sound waves and MHD turbulence are
\begin{align}
    &S_{sw}(f)= \left(\frac{f}{f_{sw}}\right)^{3}\bigg(\frac{7}{4+3\left(\frac{f}{f_{sw}}\right)^2}\bigg)^{\rfrac{7}{2}},\\
    &S_{tu}(f)=\frac{\left(\frac{f}{f_{tu}}\right)^3}{(1+\frac{f}{f_{tu}})^{\rfrac{11}{3}}(1+\frac{8\pi f}{h_{\ast}})},
\end{align}
where the red-shifted Hubble frequency $h_{\ast}$ is
\begin{equation}
    h_{\ast}=1.1\times10^{-8}\left(\frac{T_{\ast}}{100}\right)\bigg(\frac{g_{\ast}}{10}\bigg)^{\rfrac{1}{6}}.
\end{equation}
The peak frequency of difference processes are
\begin{align}
    &f_{sw}=1.3\times10^{-8}\left(\frac{1}{v_b}\right)\left(\frac{\tau}{H_{\ast}}\right)\left(\frac{T_{\ast}}{100}\right)\bigg(\frac{g_{\ast}}{10}\bigg)^{\rfrac{1}{6}},\\
    &f_{tu}=1.8\times10^{-8}\left(\frac{1}{v_b}\right)\left(\frac{\tau}{H_{\ast}}\right)\left(\frac{T_{\ast}}{100}\right)\bigg(\frac{g_{\ast}}{10}\bigg)^{\rfrac{1}{6}}.
\end{align}
The spectral shape of the GW is scaled as $f^3$ with the frequencies below the peak frequency and as $f^{-4}$ from sound waves and $f^{-5/3}$ from MHD for larger frequencies. For relativistic velocities, i.e. the runaway case, $\kappa_{v}$ is given by \cite{1004.4187},
\begin{equation}
    \kappa_{v}=\frac{\alpha_{\infty}}{0.73+0.083\sqrt{\alpha_{\infty}}+\alpha_{\infty}},
\end{equation}
and $\kappa_{tu} = \epsilon\kappa_{v}$ is the relation of the fraction of the latent heat in two processes. We take $\epsilon \simeq 0.05 $. For non-relativistic velocities, the bubble wall velocity terminates at a certain velocity slower than the speed of light. If a first-order PT take place, the transition into the true vacuum proceeds due to bubble nucleation and percolation. Dynamics of these bubbles plays a importance role in the GWs production. There are two sorts of combustion modes, detonation and deflagration. When the bubble wall expands faster than the speed of sound, combustion occurs through detonation and deflagration serves for the subsonic velocity.

In the detonation case, the Chapman-Jouguet condition \cite{Steinhardt1982, 9310044,1004.4187} gives
\begin{equation}
    \kappa_{v}=\frac{\sqrt{\alpha}}{0.135+\sqrt{0.98+\alpha}}, ~v_{b}=\frac{\sqrt{\alpha \left(\alpha+\frac{2}{3}\right)}+\sqrt{\frac{1}{3}}}{1+\alpha}.
\end{equation}While in the deflagration case,
\begin{equation}
    \kappa_{v}=\frac{6.9~v_b^{1.2}~\alpha }{1.36-0.037\sqrt{\alpha}+\alpha}  ,
\end{equation}with $v_b \le c_s$.

\subsection{Gravitational Waves Spectrum}
In the case of runaway, we need to consider all three sources of the GW. The total energy density of the GW is
\begin{equation}
    h^2 \Omega_{tot}(f)=h^2 \Omega_{en}+h^2 \Omega_{sw}+h^2 \Omega_{tu},
\end{equation}
Concerning the non-relativistic bubbles wall velocity, the GW from bubbles collision can be neglected. Therefore, the energy density only includes the contributions from the sound waves and the MHD turbulence,
\begin{equation}
h^2 \Omega_{tot}(f)=h^2 \Omega_{sw}+h^2 \Omega_{tu}.
\end{equation}
We list a short summary in each conditions as follows.
 \begin{enumerate}
   \item Runaway bubbles ($v_b \simeq c = 1$), $\alpha_\infty < \alpha$: $h^2 \Omega_{tot}(f)=h^2 \Omega_{en}+h^2 \Omega_{sw}+h^2 \Omega_{tu}$
   \item No-runaway bubbles ($v_b < c$), $\alpha_\infty > \alpha$: $h^2 \Omega_{tot}(f)=h^2 \Omega_{sw}+h^2 \Omega_{tu}$
   \begin{enumerate}
     \item detonations, $v_b > c_s$
     \item defragrations, $v_b \leq c_s$
   \end{enumerate}
 \end{enumerate}

\begin{figure}[t]
\begin{center}
\includegraphics[
height=2in, width=3.2in]
{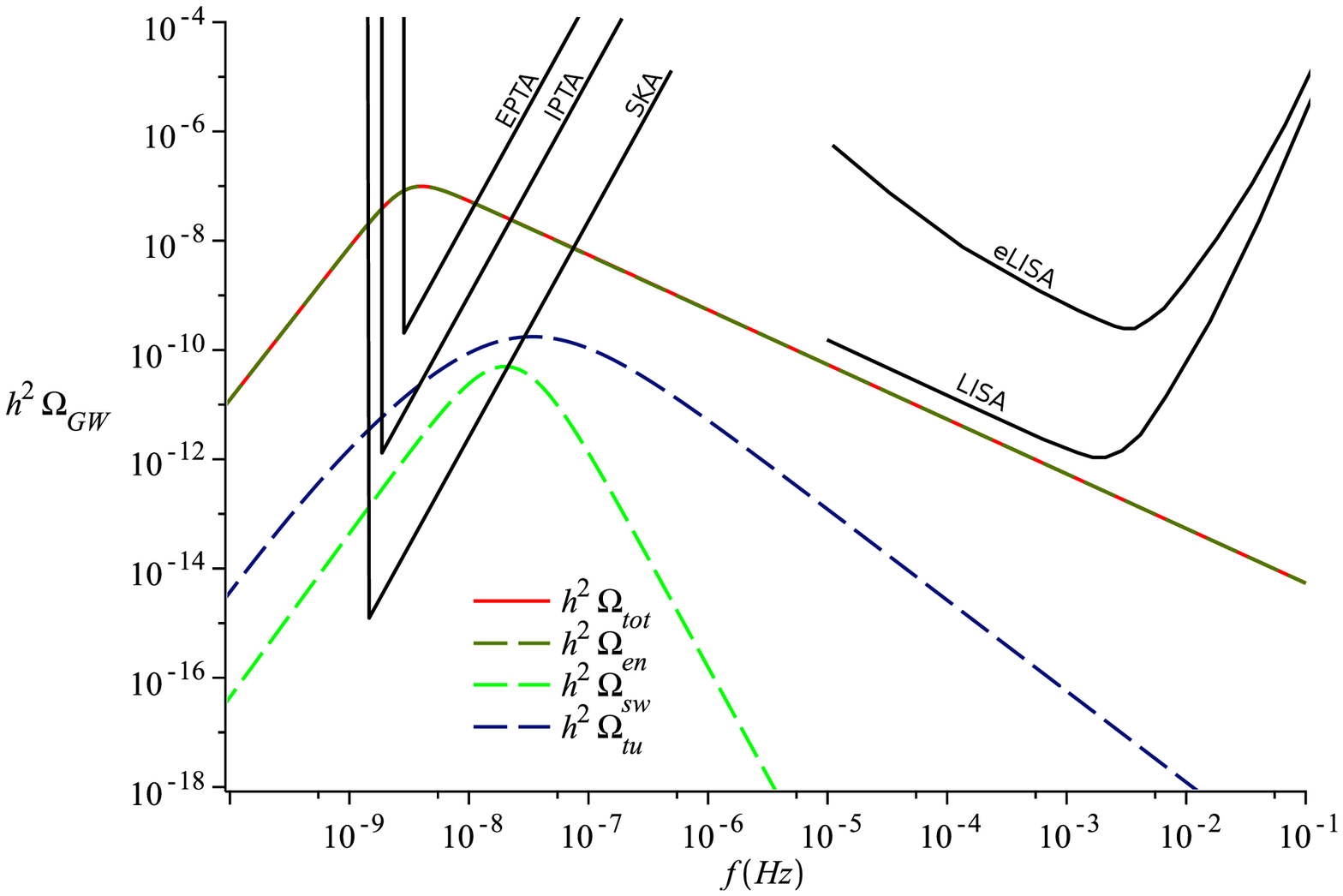} \hspace*{0.2cm} \includegraphics[
height=2in, width=3.2in]
{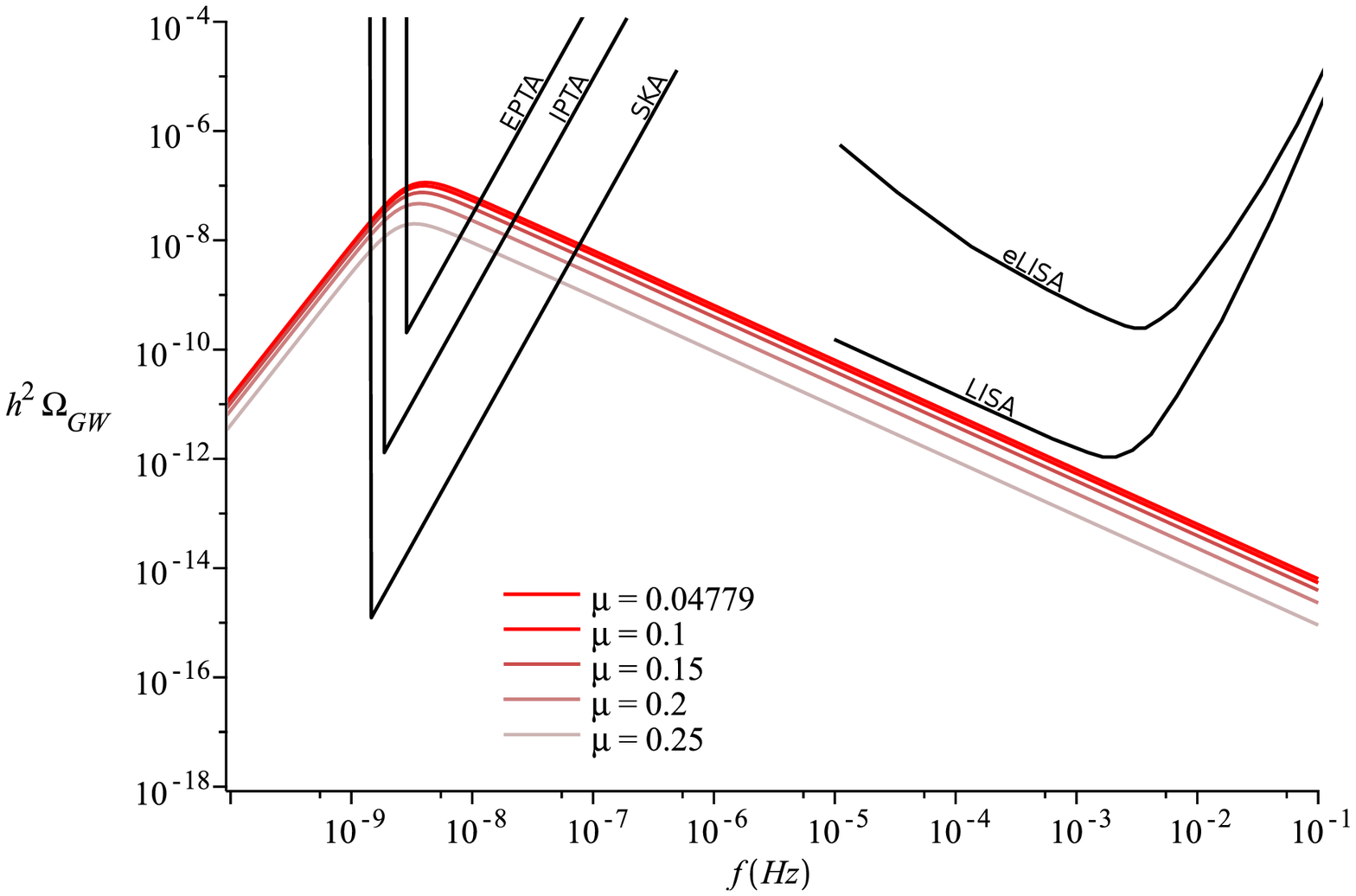} \vskip -6mm \hskip 0.15 cm \textbf{(a)} \hskip 8.3 cm \textbf{(b)}\\
\includegraphics[
height=2in, width=3.2in]
{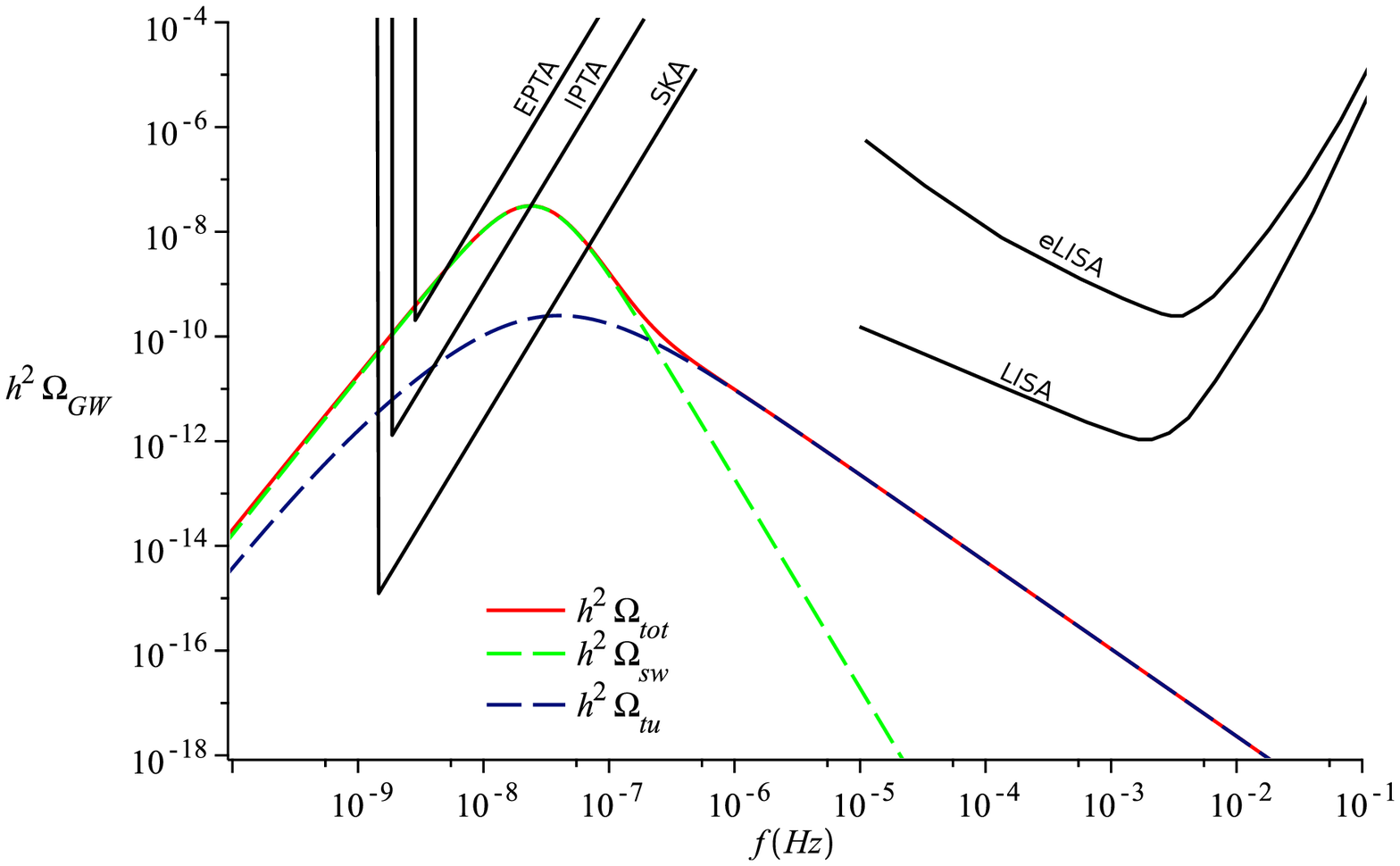} \hspace*{0.2cm} \includegraphics[
height=2in, width=3.2in]
{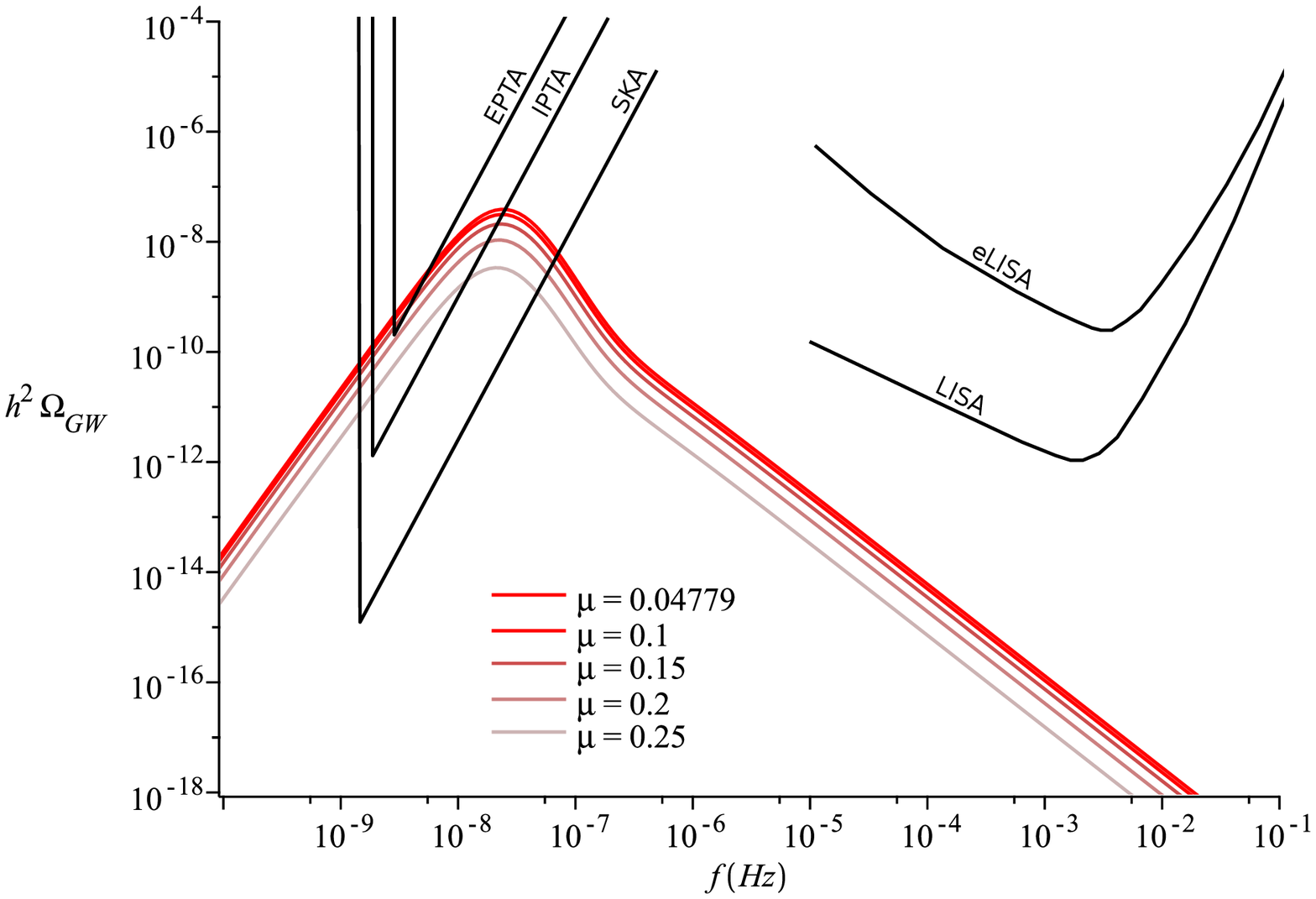} \vskip -6mm \hskip 0.15 cm \textbf{(c)} \hskip 8.3 cm \textbf{(d)}\\
\includegraphics[
height=2in, width=3.2in]
{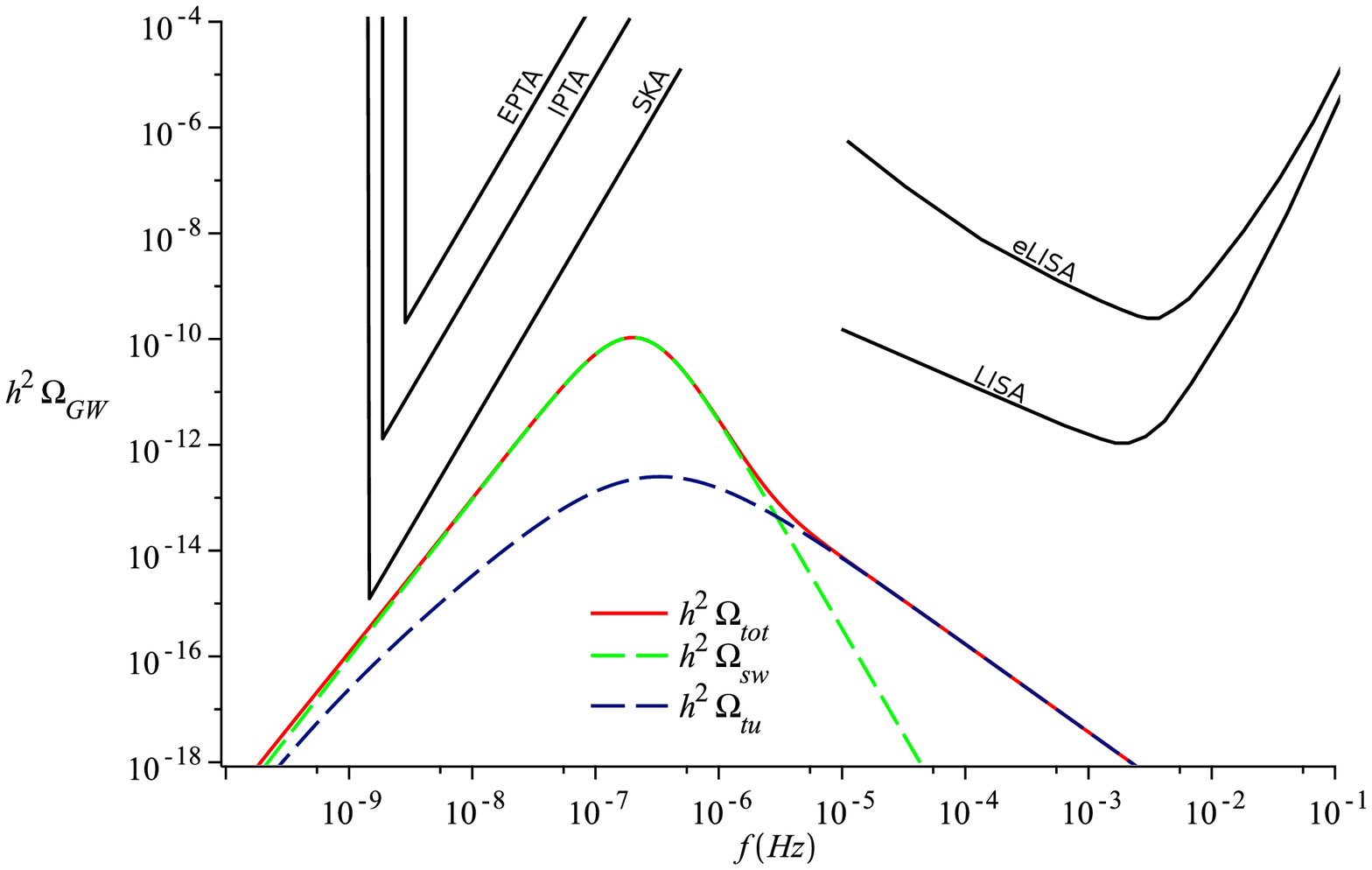} \hspace*{0.2cm} \includegraphics[
height=2in, width=3.2in]
{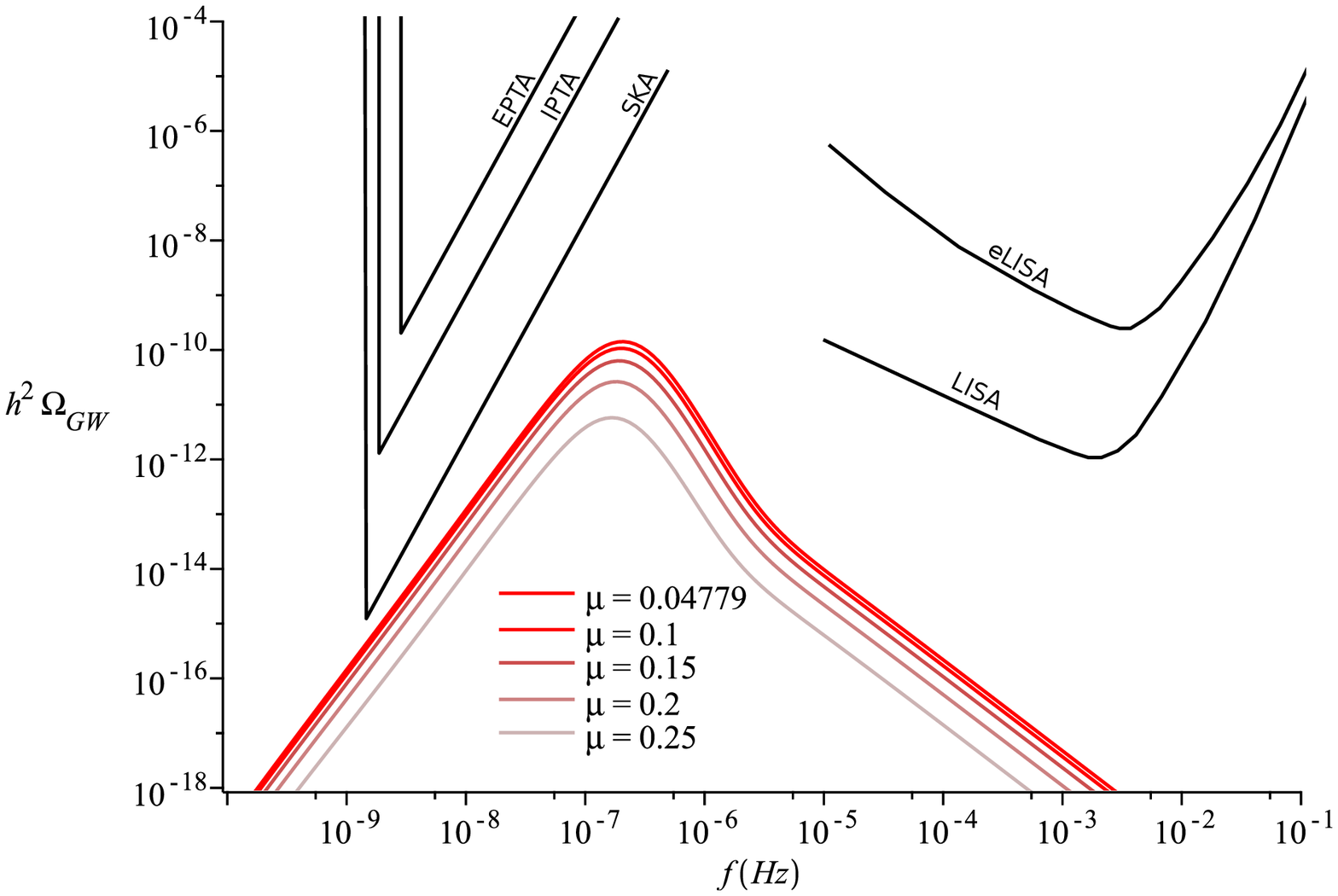} \vskip -6mm \hskip 0.15 cm \textbf{(e)} \hskip 8.3 cm \textbf{(f)}
\end{center}
\caption{For light quarks. (a)(b) The GW spectrum for $\mu=0.1 GeV$  in the runaway case with the bubble wall velocity $v_{b}=c$ and different chemical potential $\mu$. (c)(d) The GW spectrum for $\mu=0.1 GeV$ in the detonation case and different chemical potential $\mu$. (e)(f) The GW spectrum for $\mu=0.1 GeV$ in the deflagration case ($v_{b}=0.1c$) and different chemical potential $\mu$.} \label{light_spect}
\end{figure}%

\begin{figure}[t]
\begin{center}
\includegraphics[
height=2in, width=3.2in]
{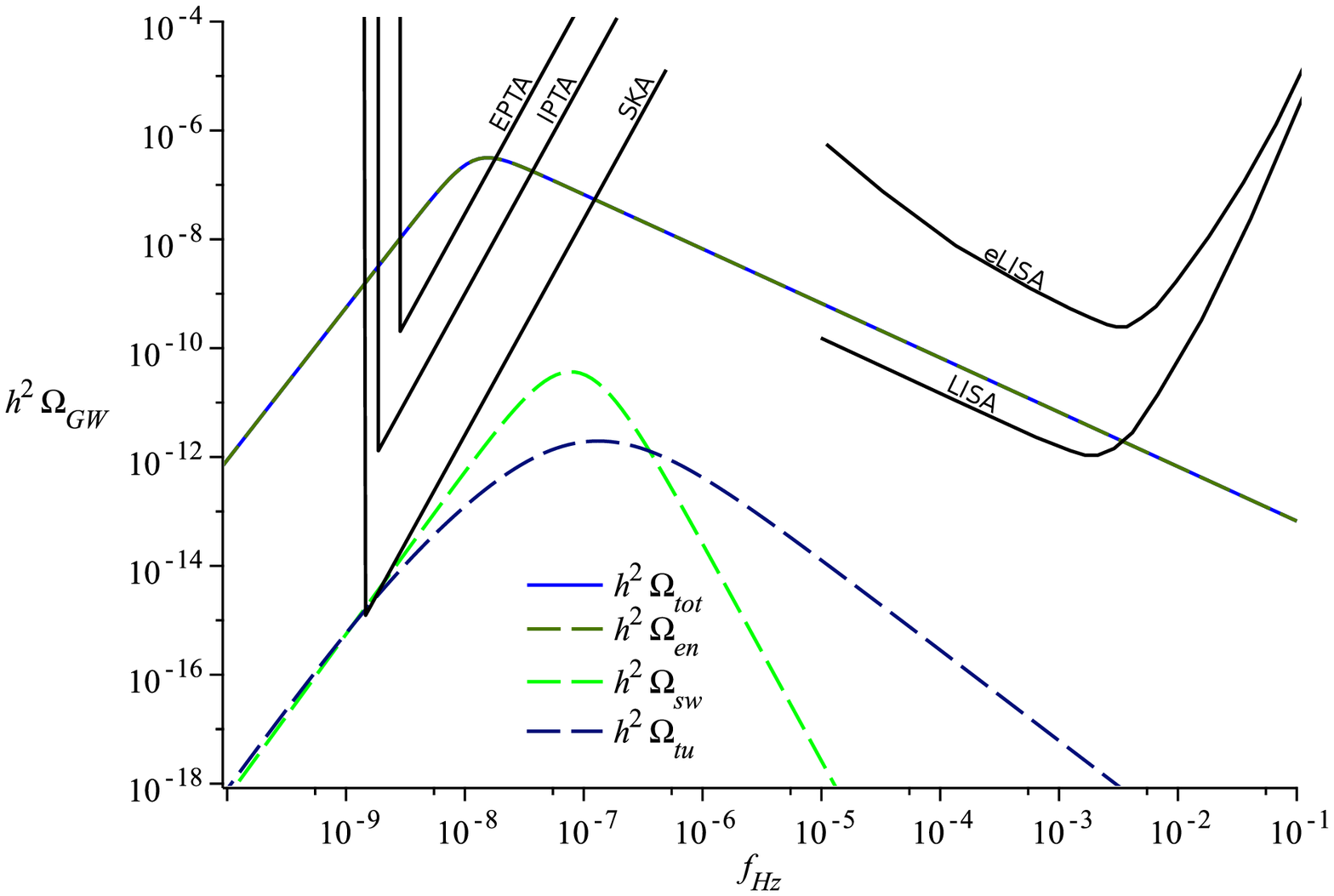} \hspace*{0.2cm} \includegraphics[
height=2in, width=3.2in]
{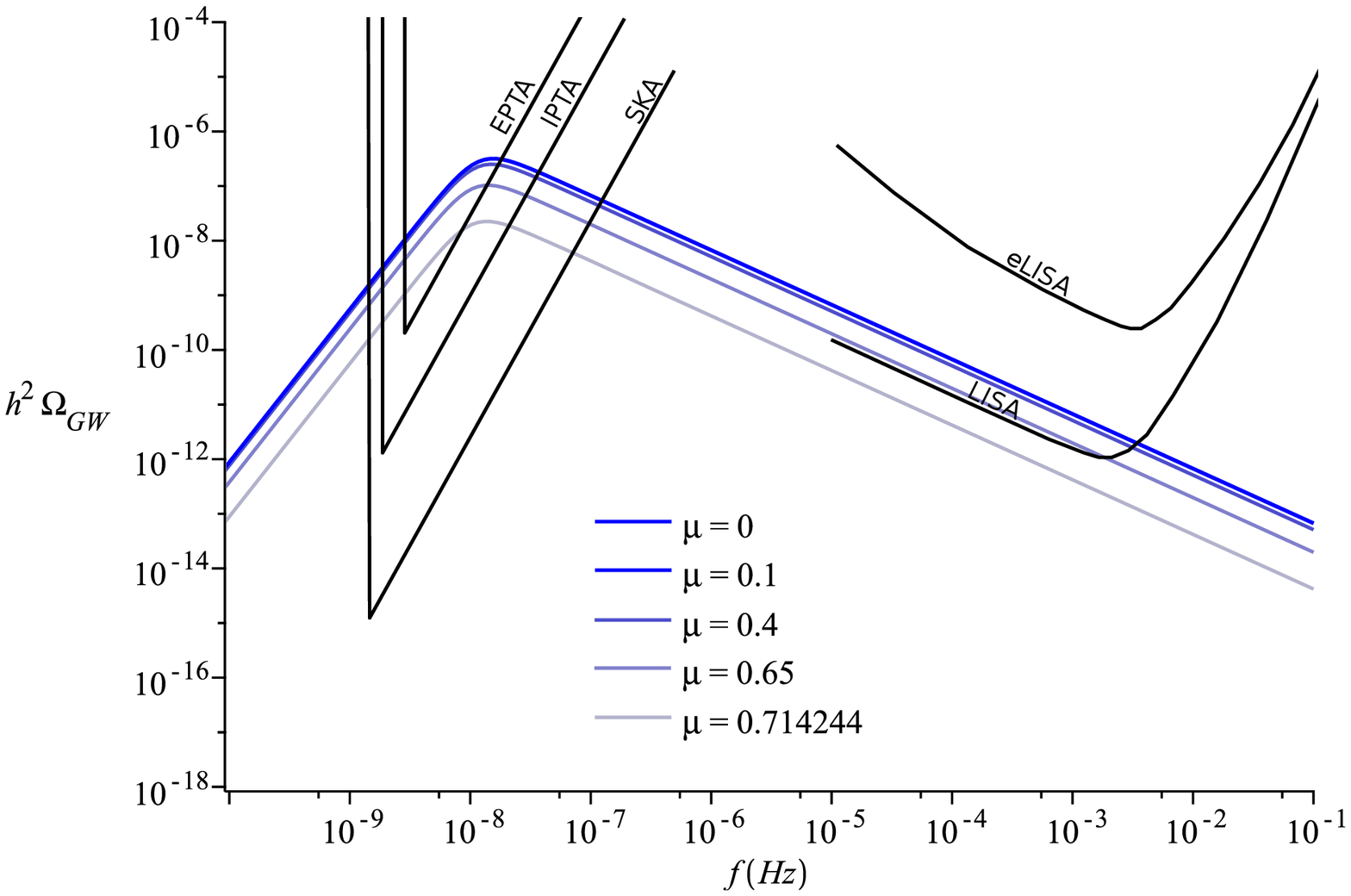} \vskip -5mm \hskip 0.15 cm \textbf{(a)} \hskip 8.3 cm \textbf{(b)}\\
\includegraphics[
height=2in, width=3.2in]
{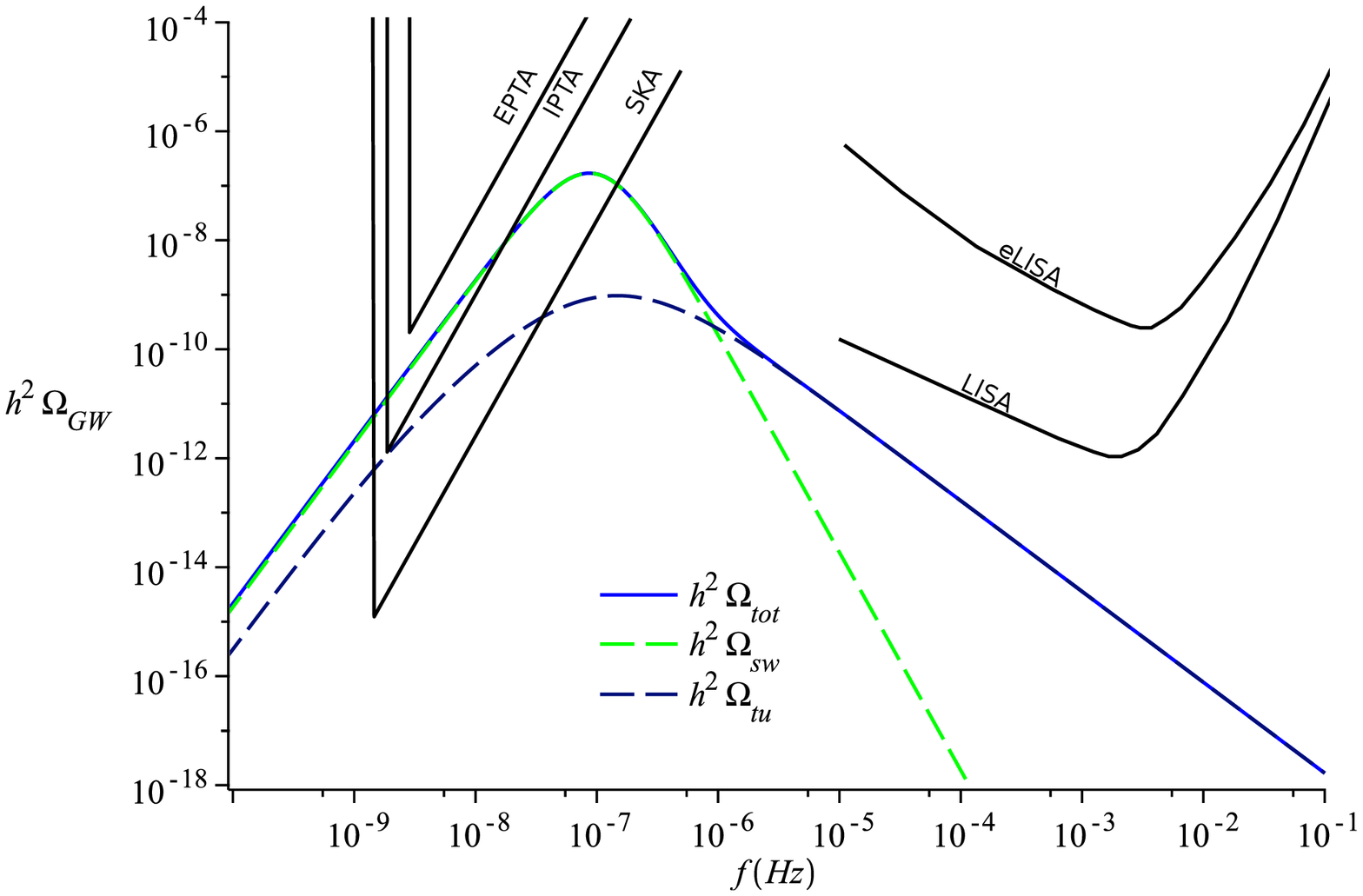} \hspace*{0.2cm} \includegraphics[
height=2in, width=3.2in]
{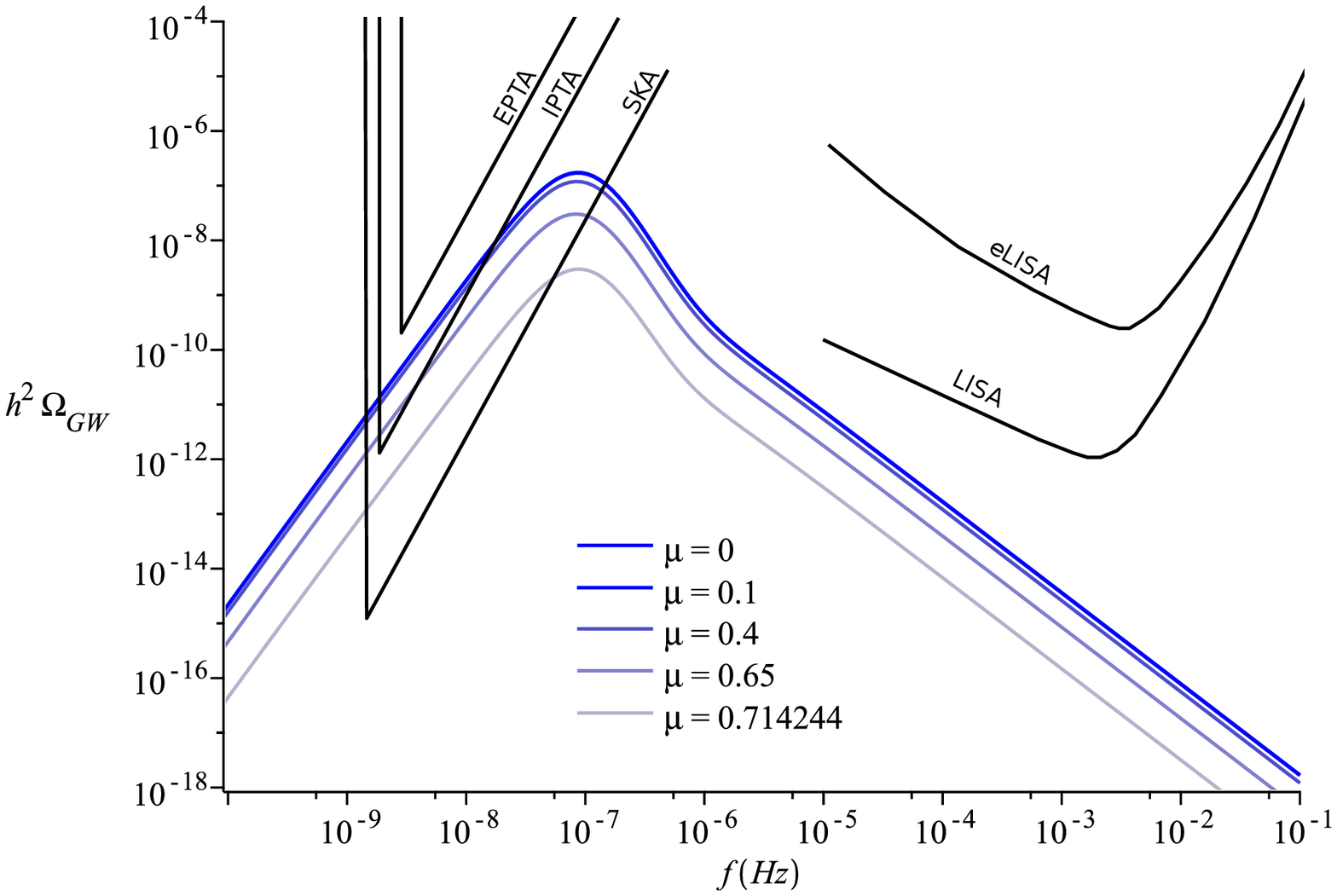} \vskip -5mm \hskip 0.15 cm \textbf{(c)} \hskip 8.3 cm \textbf{(d)}\\
\includegraphics[
height=2in, width=3.2in]
{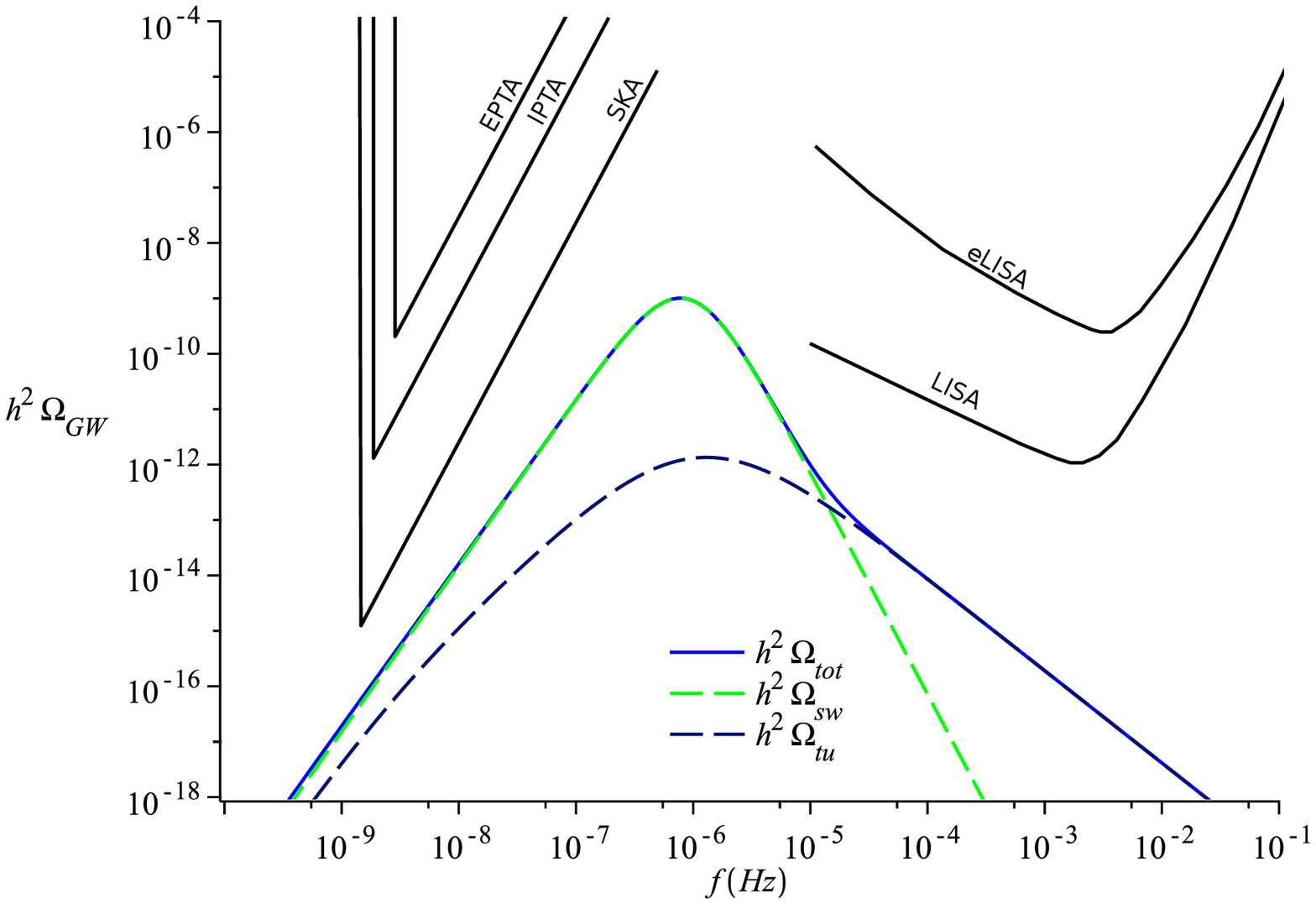} \hspace*{0.2cm} \includegraphics[
height=2in, width=3.2in]
{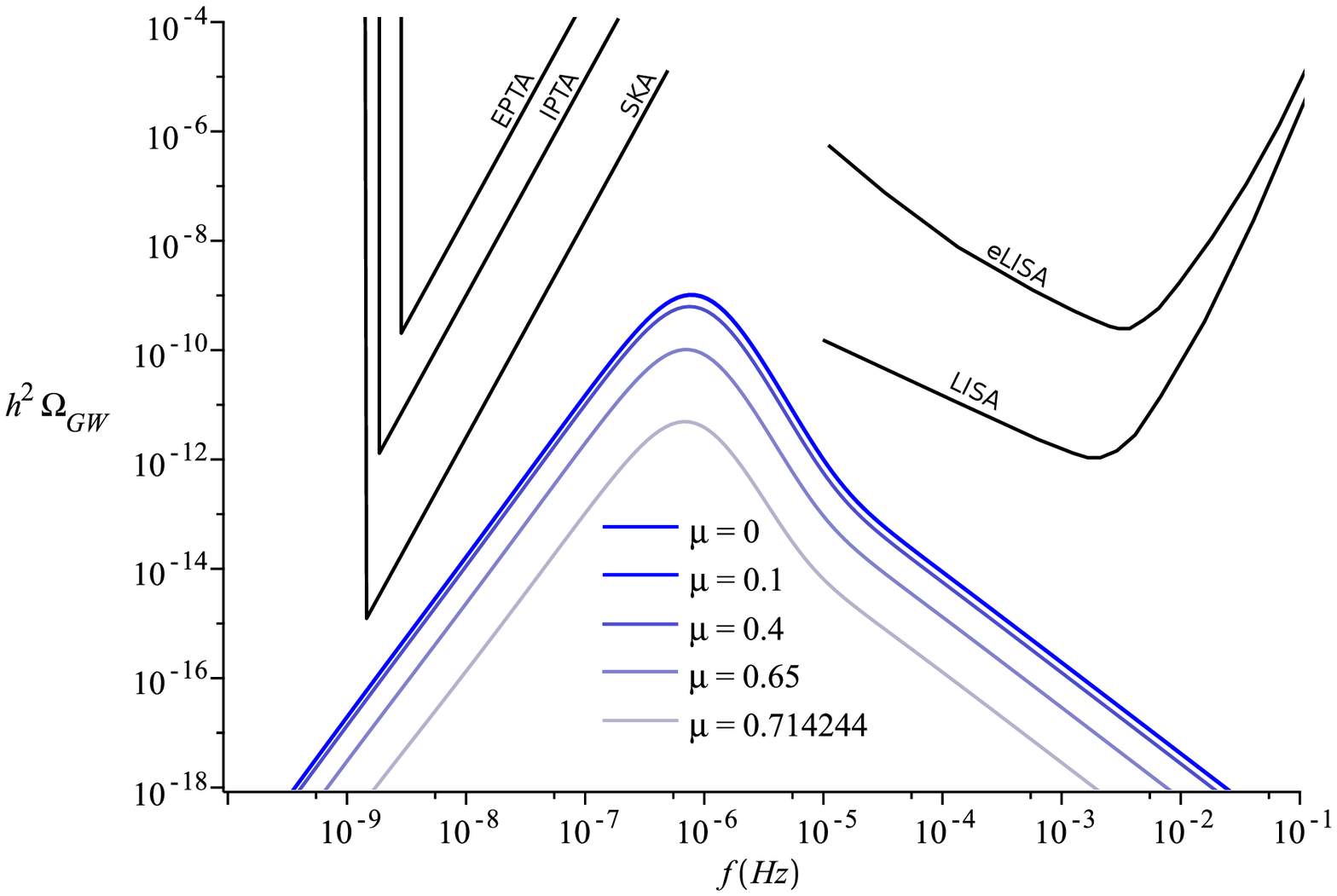} \vskip -5mm \hskip 0.15 cm \textbf{(e)} \hskip 8.3 cm \textbf{(f)}
\end{center}
\caption{For heavy quarks. (a)(b) The GW spectrum for $\mu=0.1 GeV$  in the runaway case with bubble wall velocity $v_{b}=c$ and different chemical potential $\mu$. (c)(d) The GW spectrum for $\mu=0.1 GeV$ in the detonation case and different chemical potential $\mu$. (e)(f) The GWs spectrum for $\mu=0.1 GeV$ in the deflagration case ($v_{b}=0.1c$) and different chemical potential $\mu$.} \label{heavy_spect}
\end{figure}%

The GWs spectrum for the light quarks are plotted in Fig.\ref{light_spect}. Fig.\ref{light_spect}a shows the GWs spectrum in the runaway case with the bubble wall velocity $v_{b}=c $ and the chemical potential $\mu=0.1 GeV$. All three sources contribute to the GW spectrum in this case. The dark-green dashed line represents the source of the bubble collision, the light-green dashed line represents the source of sound waves and the blue dashed line represents the source of the MHD. The total GW spectrum is plotted in the red solid line, which is dominated by the contribution from the bubble collision by about three orders. The peak of the total GW spectrum is around $10^{-8}\sim 10^{-9} Hz$ with the maximal energy density $\sim 10^{-7} GeV$, which is detectable for EPTA/IPTA/SKA. At the higher frequency, the energy density decreases quickly so that it cannot be detected by LISA/eLISA. The total GW spectrum in the runaway case with different chemical potentials are plotted in Fig.\ref{light_spect}b. Since the confinement-deconfinement transformation for the light quark near $\mu\sim 0$ is a crossover instead of a PT, we plot the total GW spectrum in Fig.\ref{light_spect}b from the critical value of the chemical potential $\mu_c=0.04779 GeV$ up to $\mu=0.25 GeV$ from top to bottom. The results illustrate that the peak of the total GW spectrum in the runaway case is almost not affected by the changing of the chemical potential, while the maximal energy density will be depressed with increasing chemical potentials that is in consistent with the results in \cite{1808.06188}. Nevertheless, the total GW spectrum in the runaway case is always in the detectable scope of EPTA/IPTA/SKA.

Fig.\ref{light_spect}c and Fig.\ref{light_spect}e shows the GWs spectrum in the non-runaway cases with the chemical potential $\mu=0.1 GeV$. Fig.\ref{light_spect}c is for the denotation case with the bubble wall velocity satisfying the Chapman-Jouguet condition. Fig.\ref{light_spect}e is for the deflagration case with the bubble wall velocity $v_{b}=0.1c$. In the both non-runaway cases, only two sources contribute to the GW spectrum. The light-green dashed line represents the source of sound waves and the blue dashed line represents the source of the MHD. The total GW spectrum is plotted in the red solid line, which is dominated by the contribution from the sound waves at low frequency and is dominated by the contribution from the MHD at high frequency. The peaks of the total GW spectrum are around $10^{-8} Hz$ in the denotation case and around $10^{-7} Hz$ in the deflagration case, respectively. The maximal energy density in the denotation case $\sim 10^{-8} GeV$, which is in the detectable scope of EPTA/IPTA/SKA. The maximal energy density in the deflagration case $\sim 10^{-11} GeV$ and is not detectable in any of the proposed detectors. At the higher frequency, similar to the runaway case, the energy densities decreases quickly so that they cannot be detected by LISA/eLISA. The total GW spectrum in the non-runaway cases with different chemical potentials are plotted in Fig.\ref{light_spect}d and Fig.\ref{light_spect}f. The results illustrate that the peak of the total GW spectrum in the non-runaway cases are almost not affected by the changing of the chemical potential, while the maximal energy density will be depressed with increasing chemical potentials \cite{1808.06188}. The total GW spectrum in the detonation case is detectable up to $\mu=0.25 GeV$ for SKA.

The GW spectrum for the heavy quarks are plotted in Fig.\ref{heavy_spect}. The behaviors of the GW spectrum for the heavy quarks are similar to that for the light quarks, but with higher energy density. In the runaway case, the GW spectrum is detectable for not only EPTA/IPTA/SKA but also for LISA at higher frequency. In the detonation case, the GW spectrum is detectable for both IPTA and SKA. However, the deflagration case is still out of the detectable scope of any proposed detector.

\begin{figure}[t!]
\begin{center}
\includegraphics[
height=4.2in, width=6.6in]
{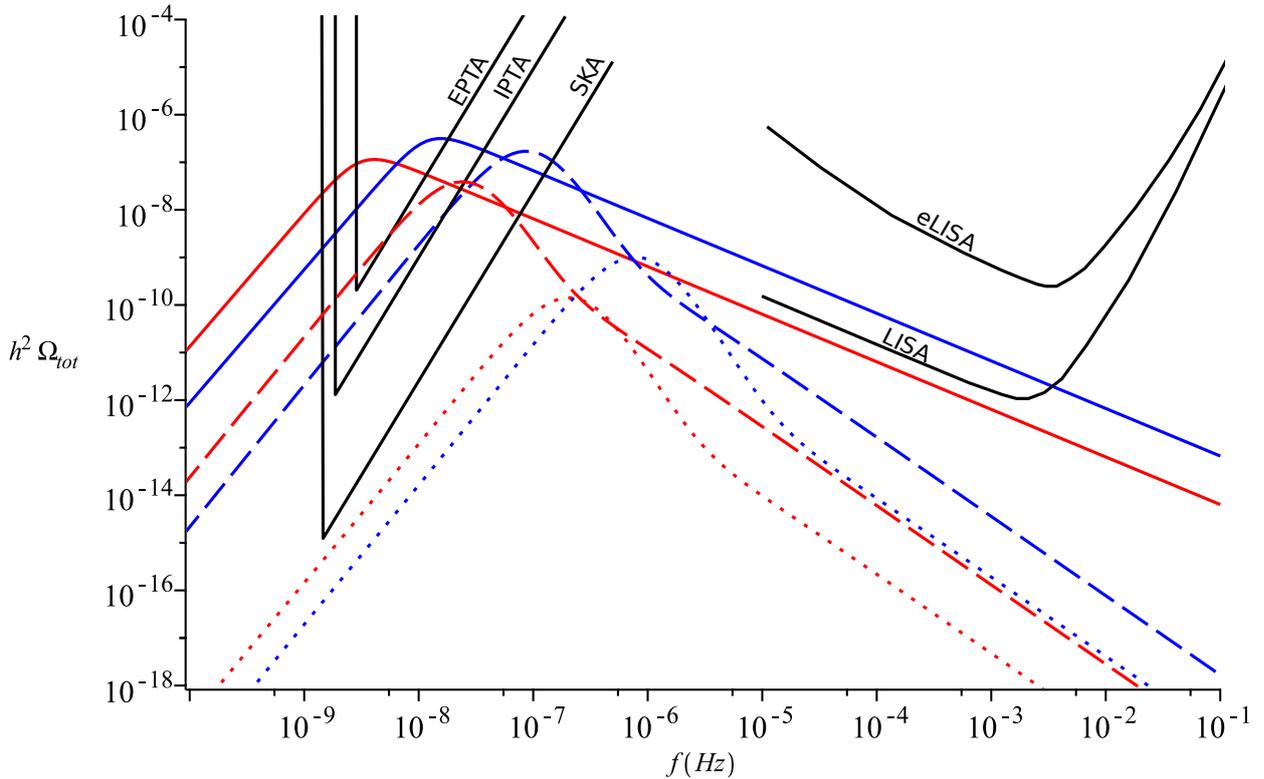}
\end{center}
\caption{GW spectrum for heavy (blue) and light (red) meson at chemical potential $\mu=0.1 GeV$ in different sectors: runaway case is solid line, detonation case is dash line and deflagration is point line. It is clear that the transition scale between heavy/light meson confinement PT also perform in the gravitational spectrum.} \label{fig_casecombine}
\end{figure}%
 We summary our results in Fig.\ref{fig_casecombine}, in which the total GW spectrum in the runaway, detonation and deflagration cases for both light and heavy quarks are plotted. The red lines are for the light quarks, while the blue lines are for the heavy quarks. Furthermore, the solid/dashed/dotted lines represent the total GW spectrum in the runaway/detonation/deflagration cases. In the runaway case, the GW spectrum for both light and heavy quarks are detectable for EPTA/IPTA/SKA. In addition, the GW spectrum for heavy quarks is detectable for LISA. In the detonation case, the GW spectrum for heavy quarks can be detected by SKA and maybe IPTA, while that for light quarks just passes the noise curve of SKA a little bit and could be detected if lucky enough. But at high frequency, both of them are not detectable for LISA/eLISA. In the deflagration case, the GW spectrum for either light or heavy quarks are not detectable for the currently proposed detectors. To sum up, the GW spectrum in the runaway case has much more chance to be detected than the other two cases, especially for heavy quarks.

\begin{figure}[t!]
\begin{center}
\includegraphics[
height=1.8in, width=2.22in]
{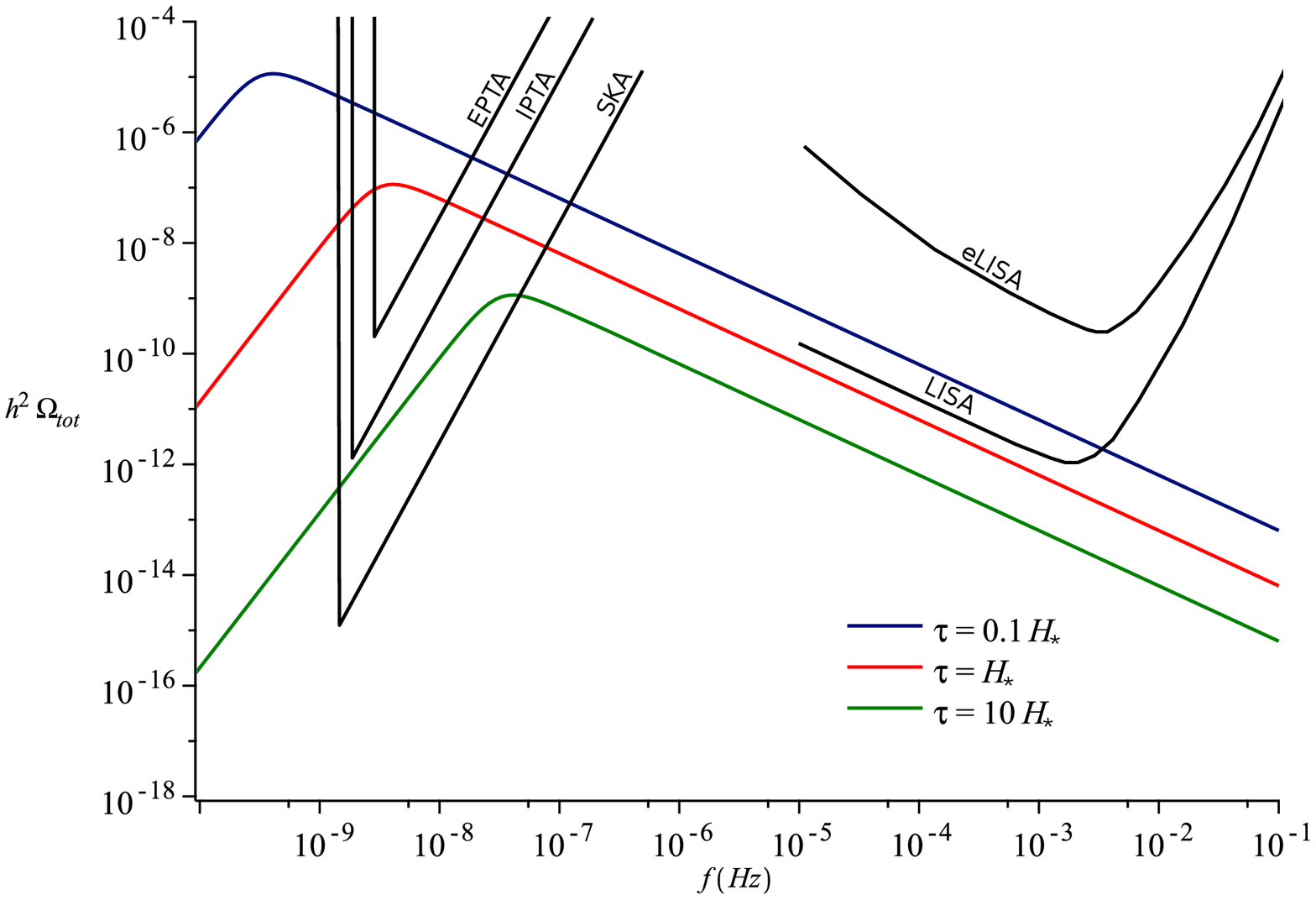} \hspace*{0.07cm} \includegraphics[
height=1.8in, width=2.22in]
{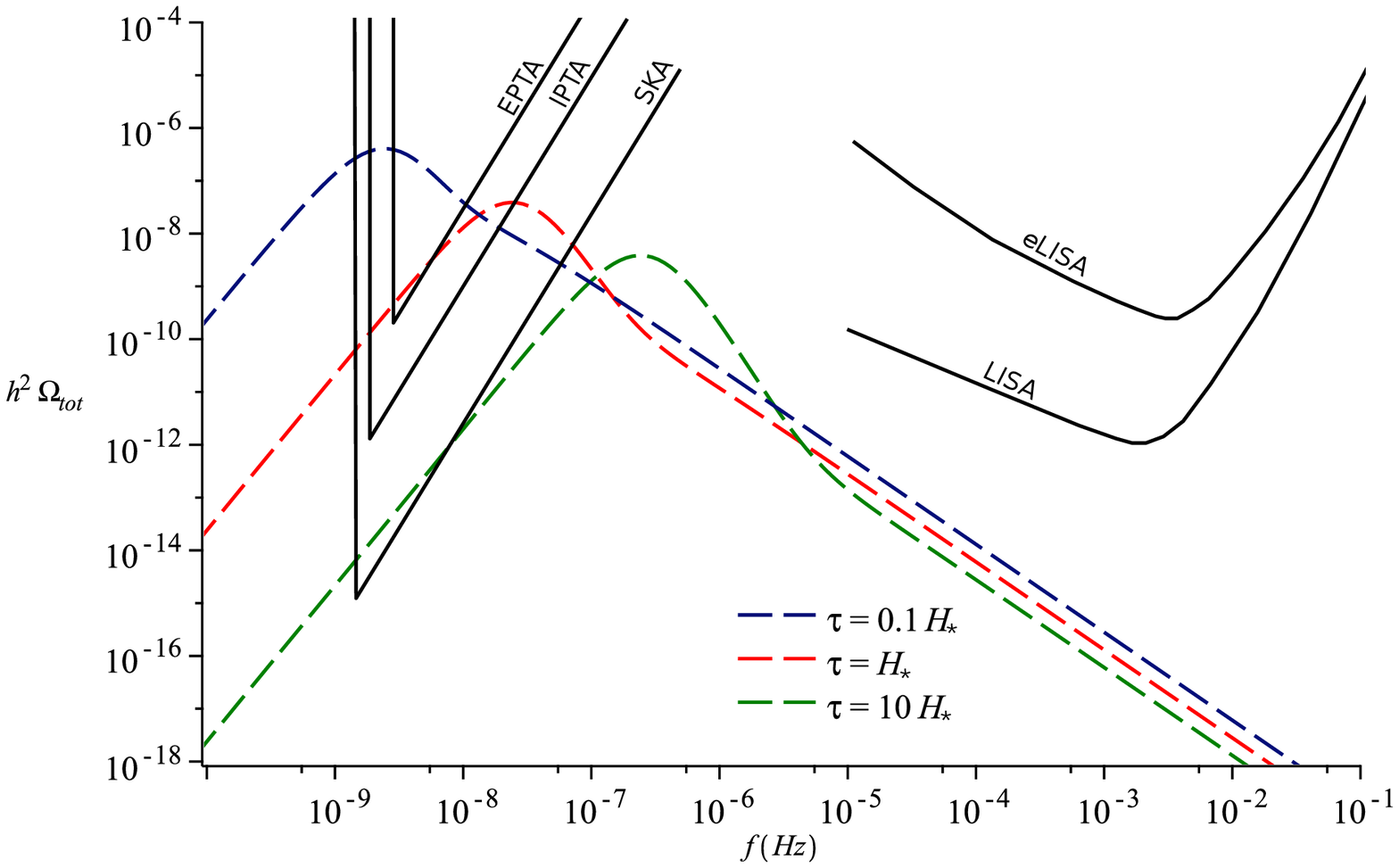} \hspace*{0.07cm}
\includegraphics[
height=1.8in, width=2.22in]
{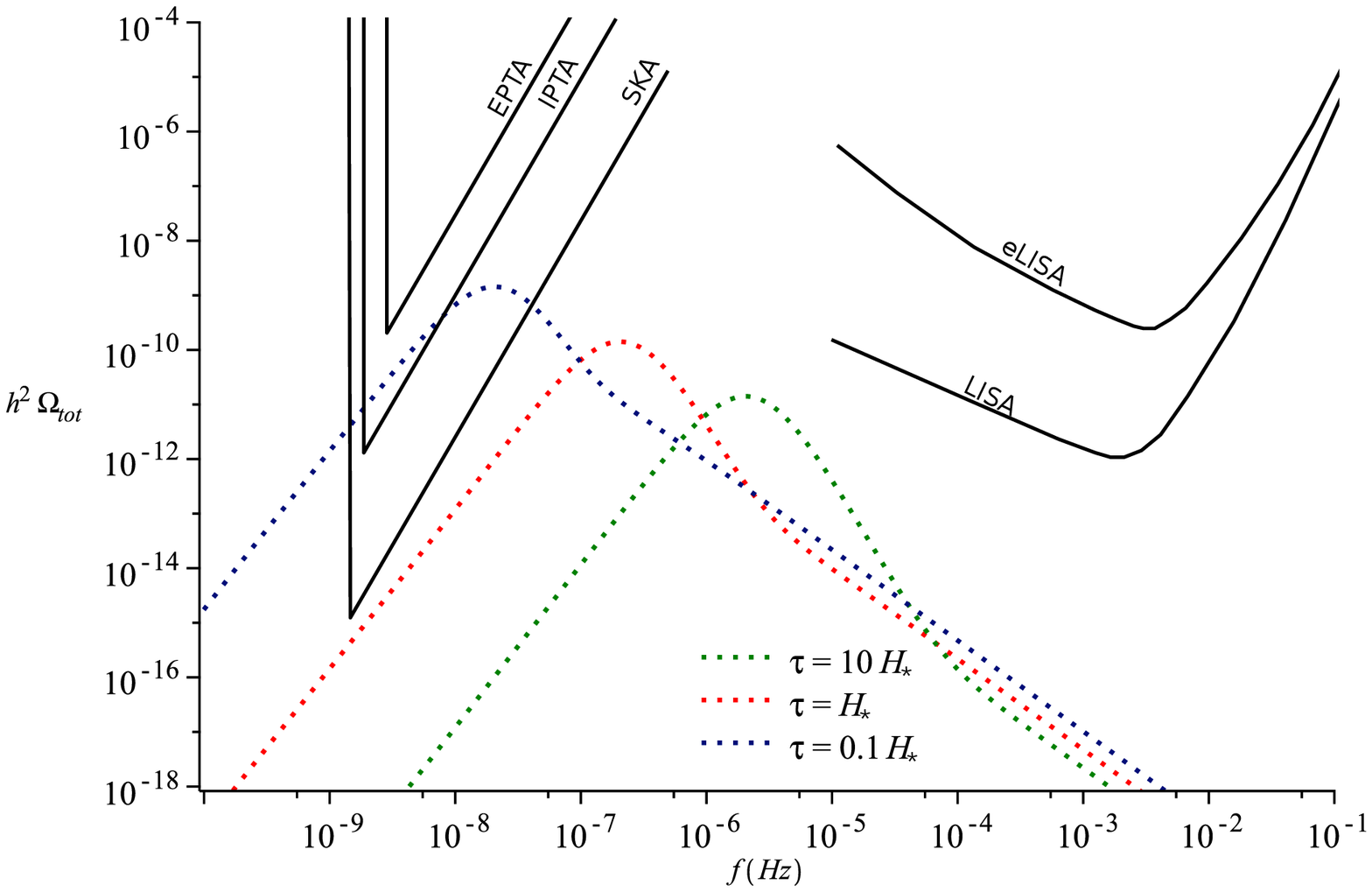}
\vskip -0.05cm \hskip 0.15 cm \textbf{(a)} \hskip 5.4 cm \textbf{(b)}\hskip 5.4 cm\textbf{(c)}
\end{center}
\caption{The GW spectrum (a), (b) and (c) are calculated from light quark model with different cases at chemical potential $\mu=0.1 GeV$ in different duration $\tau^{-1}$.} \label{fig_difftau}
\end{figure}

In the most of this paper, we have fixed the duration of the PT $\tau=H_*$. However, the reasonable range for the duration of the PT is around $0.1H_{*} \sim 10 H_{*}$. The different value of the duration will affect the behavior of the GW spectrum. We demonstrated the GW spectrum of light quarks with $\mu=0.1 GeV$ for different duration. The results are plotted in Fig.\ref{fig_difftau}. The blue, red and green lines represent the GW spectrum with the duration  $\tau=0.1H_*, H_*, 10H_*$, respectively. The results shows that reducing the duration of the PT significantly enhances the energy density of the GW spectrum, and eventually increases the possibility to detect the GW by the currently proposed detectors. In the runaway case, the GW spectrum with  $\tau=0.1H_*$ becomes detectable for LISA. In the detonation case, the GW spectrum with  $\tau=0.1H_*$ is clearly detectable for EPTA/IPTA/SKA. In the deflagration case, the GW spectrum with $\tau=H_*$ could not be detected by any proposed detector, but with $\tau=H_*$, it is detectable for IPTA/SKA.

\section{Conclusion}
In this paper, we have studied the GW spectrum created from the first-order PT during the expansion of the early universe. The PT is identified with the confinement-deconfinement transition in QCD. We considered two holographic QCD models for either light or heavy quarks, respectively.

The energy density of the GW spectrum from various sources: bubble collision, sound waves and MHD turbulence have been calculated by using the numerical fitting formula. Classified by the bubbles wall velocity $v_b$, there are three cases: runaway ($v_b\sim c$), detonations ($c_s<v_b<c$) and defragrations  ($v_b\le c_s$). We calculated the GW spectrum for both light and heavy quarks in all three cases with different chemical potentials and duration of the PT. We compared the energy density of the GW spectrum with the noise curves of the currently proposed GW detectors: LISA/eLISA for the frequency $10^{-1} \sim 10^{-5} Hz$ and EPTA/IPTA/SKA for the frequency $10^{-6} \sim 10^{-9} Hz$.\\

We list our results as follows:
\begin{itemize}
\item In the runaway case, the dominant contribution comes from the bubble collision.
\item In the detonation and deflagration cases, the effect of the bubble collision can be ignored, and the dominant contribution comes from the sound waves for the low frequency and the MHD turbulence for the high frequency.
\item Increasing the chemical potential decreases the energy density of the GW spectrum.
\item Reducing the duration of the phase transition significantly enhances the energy density of the GW spectrum.
\item For light quarks with $\mu=0.1 GeV$ and $\tau=H_*$,
   \begin{enumerate}
     \item in the runaway case, the GW spectrum is detectable by EPTA/IPTA/SKA,
     \item in the denotation case, the GW spectrum is detectable by IPTA/SKA, and might be by EPTA,
\item in the deflagration case, the GW spectrum is not detectable at all by the currently proposed detectors.
   \end{enumerate}
\item For heavy quarks with $\mu=0.1 GeV$ and $\tau=H_*$,
   \begin{enumerate}
     \item in the runaway case, the GW spectrum is detectable by EPTA/IPTA/SKA, as well as LISA,
     \item Iin the denotation case, the GW spectrum is detectable by IPTA/SKA,
\item in the deflagration case, the GW spectrum is not detectable at all by the currently proposed detectors.
   \end{enumerate}
 \end{itemize}

\subsection*{Acknowledgements}
This work is supported by the Ministry of Science and Technology (MOST 106-2112-M-009-005-MY3) and in part by National Center for Theoretical Science (NCTS), Taiwan.

\end{document}